\documentclass[dvipdfm,aps,prd,showpacs,showkeys,preprintnumbers,amsmath,amssymb,nofootinbib,11pt]{revtex4}
\usepackage{eepic}
\usepackage{indentfirst}
\usepackage{mathrsfs}
\usepackage{fancyhdr}
 \usepackage{ulem}
\usepackage{tabularx}
\usepackage{float}
\usepackage{graphicx}
\usepackage{slashed}
\usepackage[colorlinks]{hyperref}

\newcommand{\be}{\begin{equation}}
\newcommand{\ee}{\end{equation}}
\newcommand{\ba}{\begin{array}{c}}
\newcommand{\ea}{\end{array}}
\newcommand{\bqa}{\begin{eqnarray}}
\newcommand{\eqa}{\end{eqnarray}}
\newcommand{\bqaa}{\begin{eqnarray*}}
\newcommand{\eqaa}{\end{eqnarray*}}
\newcommand{\mL}{\mathcal{L}}

\newcommand{\bra}{\langle}
\newcommand{\ket}{\rangle}
\newcommand{\nn}{\nonumber}

\begin{document}

\title { Unified study of $J/\psi \to PV$, $P\gamma^{(*)}$ and light hadron radiative processes }

\author{ Yun-Hua~Chen$^a$}\thanks{chenyh@ihep.ac.cn}
\author{ Zhi-Hui~Guo$^{b,c}$}\thanks{ Corresponding author: zhguo@mail.hebtu.edu.cn }
\author{ Bing-Song~Zou$^{a,c}$}\thanks{zoubs@itp.ac.cn}
\affiliation{ ${}^a$  Institute of High Energy Physics, CAS, Beijing
100049, People's Republic of China
 \\ ${}^b$ Department of Physics, Hebei Normal University, Shijiazhuang 050024, People's Republic of China
  \\ ${}^c$ State Key Laboratory of Theoretical Physics, Institute of
Theoretical Physics, CAS, Beijing 100190, People's Republic of China
}

\begin{abstract}

Within the framework of the effective Lagrangian approach, we perform a
thorough analysis of the $J/\psi \to P\gamma(\gamma^*)$, $J/\psi \to
VP$, $V\to P\gamma(\gamma^*)$, $P\to V\gamma(\gamma^*)$ and $P\to\gamma\gamma(\gamma^*)$ processes,
where $V$ stand for light vector resonances, $P$ stand for light
pseudoscalar mesons, and $\gamma^*$ subsequently decays into lepton
pairs.  The processes with light pseudoscalar mesons $\eta$ and
$\eta'$ are paid special attention to and the two-mixing-angle
scheme is employed to describe their mixing. The four mixing
parameters both in singlet-octet and quark-flavor bases are updated
in this work. We confirm that the $J/\psi \to \eta(\eta^{\prime})\gamma^{(*)}$ processes are predominantly
dominated by the $J/\psi\to \eta_c \gamma^{*} \to
\eta(\eta^{\prime})\gamma^{(*)}$ mechanism.
Predictions for the $J/\psi \to P \mu^+\mu^-$ are presented. A detailed discussion on
the interplay between electromagnetic and strong transitions in the
$J/\psi \to VP$ decays is given.

\end{abstract}


\maketitle

\section{Introduction}

The vast decay modes of the $J/\psi$ into light flavor hadrons provide us invaluable information
on the mechanisms of light hadron production from the $c\bar{c}$ annihilation and they are also ideal for
the study of light hadron dynamics, such as the $SU(3)$-flavor symmetry breaking and Okubo-Zweig-Iizuka(OZI) rules.
We focus on two types of $J/\psi$ decays in this work, i.e. $J/\psi \rightarrow PV$ and $J/\psi \rightarrow P\gamma(\gamma^*)$,
with $V$ the light vector resonances and $P$ the light pseudoscalar mesons.

For the charmonium radiative decays $J/\psi \rightarrow P\gamma$, the dominant underlying mechanism is the $c\bar{c}$
annihilation into two gluons plus a photon, as advocated in many previous
works~\cite{seiden88,Feldmann98,Thomas,Novikov,ohta80,gilman87,KTChao1990,Escribano99,Escribano05,kroll05,gerard05,QZhao2011,HBLi2012}.
While for the $J/\psi \rightarrow PV$ decays, both electromagnetic (EM) and strong interactions will enter and an important issue
is the interplay between the two parts, that has been extensively studied in 
literature~\cite{haber85,seiden88,Ball,bramon97,Gerard:1999uf,Feldmann98,QZhao2007,QZhao2008,Thomas,Escribano2010}.
All of the attempts to understand these $J/\psi \rightarrow PV$ decays are based on the similar model with slight variations.
In this model, the dominant part of the amplitude is assumed to proceed through the $c\bar{c}$ annihilation into light hadrons
via three gluons, which is the so-called the single-OZI suppressed diagram. Later on, doubly OZI suppressed diagrams are also
introduced, where an additional gluon is exchanged between the vector and pseudoscalar mesons. For the EM interaction pieces, there
are also two different kinds of diagrams, the singly disconnected one (one photon exchange) and doubly disconnected ones.
We refer to Ref.~\cite{seiden88} for a detailed discussion on different mechanisms.
Based on these arguments, the previous research works, such as those in 
Refs.~\cite{haber85,seiden88,Ball,bramon97,Feldmann98,Gerard:1999uf,QZhao2007,QZhao2008,Thomas,Escribano2010},
proceed the discussion through directly writing down the amplitudes by introducing some phenomenological couplings for different processes.
The $SU(3)$ symmetry breaking effects are also introduced at the amplitude level.

In this work, we do not follow the previous routine to further scrutinize and  refine different terms in the amplitudes, instead we
start from the very beginning by constructing the relevant effective Lagrangians and then use them to calculate the amplitudes.
One of the advantages starting from
the effective Lagrangian approach is that it allows us to make a systematic study of different processes by simultaneously taking
into account different mechanisms in a consistent and transparent way. This approach is specially useful to incorporate the OZI rule
and $SU(3)$-flavor breaking effects.

The only theoretical framework that is generally accepted to account for the successful OZI rule in various hadronic processes
is the large $N_C$ QCD~\cite{lnc}. It has been demonstrated in Ref.~\cite{gl845} that one can build a simple relation between the number
of flavor traces in effective field theory and the $N_C$ counting rule, though some care should be paid attention to in special cases due to
the subtlety of using matrix relations among the traces of products~\cite{gl845}. Generally speaking,
to introduce one additional trace to an operator in the effective Lagrangian will make this operator one more order suppressed by $1/N_C$,
i.e. one more order of OZI suppression. Therefore, it is convenient and easy to
systematically include the OZI suppressed effects in the effective Lagrangian approach. Another important benefit to work in the effective Lagrangian framework
for the processes of $J/\psi$ decaying into light hadrons is to properly incorporate the $SU(3)$-flavor symmetry breaking effects.
In the chiral limit, QCD exhibits the strict $SU(3)_L \times SU(3)_R \to SU(3)_V$ spontaneous-symmetry-breaking pattern,
leading to eight massless pseudo Nambu-Goldstone bosons (pNGBs), which obey an exact $SU(3)$ symmetry.
This exact $SU(3)$-flavor symmetry has to be broken in order to be consistent with the small but non-vanishing masses of $\pi,K,\eta$.
The strong $SU(3)$-flavor symmetry breaking in QCD is implemented through the introduction of explicit non-vanishing quark masses.
This feature of QCD is elegantly embedded in chiral perturbation theory ($\chi$PT)~\cite{gl845} through the chiral building block operators $\chi_{\pm}$,
which we will explain in detail later.  Apparently the chiral power counting is by no means applied to the pNGBs in $J/\psi$ decays, since
the momenta of pNGBs are far beyond the validity region allowed by $\chi$PT.
Nevertheless even we do not have the chiral power counting, other ingredients from
chiral effective field theory can be still useful for us to construct the relevant effective Lagrangian for $J/\psi$ decays into light hadrons, such
as the well-established chiral building blocks incorporating the light pNGBs and the systematic
way to consider the $1/N_C$ or OZI suppressed and $SU(3)$ symmetry breaking effects.

In addition to the light pNGBs, we also need to include the dynamical fields of light vector resonances.
Guided by chiral symmetry and large $N_C$ expansion, resonance chiral theory (R$\chi$T)~\cite{Ecker} provides us
a reliable theoretical framework to study the interaction between the light flavor resonances and pNGBs in the intermediate energy region
and it has been successfully applied in many phenomenological
processes~\cite{Guo:2011pa,Guo:2012yt,Guo:2010dv,Jamins,Roig:2013baa,Roig:2014uja,chen2012,chen2014,Zhou:2014ana}.
The building blocks involving
resonance states from R$\chi$T~\cite{Ecker} will be also employed in our present work
to construct the relevant effective Lagrangian describing the interactions between light hadrons and the $J/\psi$.
These Lagrangians offer an efficient and systematic framework to analyze the processes of
$J/\psi \rightarrow P\gamma$, $J/\psi \rightarrow VP$ and $J/\psi\rightarrow Pl^+l^-$, with the leptons $l=e,\mu$.

From the experimental point of view, the first measurements of $J/\psi \to P\gamma^* \to P e^+ e^-$ ($P=\pi^0,\eta,\eta^{\prime}$)
are performed by the BESIII collaboration very recently~\cite{BESIII2014} and
updated world average results for $J/\psi \to PV$ and $P\gamma$ are available~\cite{Pdg} as well.
These new measurements and updated experimental results will be definitely useful to pin down the unknown couplings in our theoretical model and hence
to reveal the underlying mechanisms of $J/\psi$ decays into light hadrons.
For $J/\psi \rightarrow \eta e^+e^-$ and $J/\psi \rightarrow \eta^{\prime} e^+e^-$, the vector-dominant-model (VMD) predictions
of the decay rates are consistent with the experimental data. While
there are around 2.5 standard deviations between the theoretical prediction
and the measurement for $J/\psi \rightarrow \pi^0 e^+e^-$ process, which deserves further study~\cite{HBLi2012}.

Another important issue we will address in this article is the
properties of $\eta$ and $\eta^{\prime}$ mesons. The composition of
the $\eta$ and $\eta^{\prime}$ mesons has long been a subject of
theoretical discussions~\cite{Gilman,Feldmann97,Thomas,Feldmann00}
and is of current interest with many new measurements with high
statistics and high precision~\cite{kloe,jlab,besiii}. In Ref.~\cite{Leutwyler98}, the
two-mixing-angle description has been proposed to settle the
$\eta-\eta'$ mixing, going beyond the conventional one-mixing-angle
description~\cite{Feldmann00}. The robustness of the
two-mixing-angle description scheme has been confirmed in various
analyses~\cite{Feldmann98,Feldmann99,Ball,Escribano99,Pennington,Benayoun00,Goity,
Escribano01,Escribano05,pham10,chen2012}, and we have provided a mini-review
on the $\eta-\eta^{\prime}$ mixing in Ref.~\cite{chen2012}. In this
article, we extend the previous work of Ref.~\cite{chen2012} by including the $J/\psi$ decays: $J/\psi \rightarrow P\gamma$, $J/\psi \rightarrow VP$,
$J/\psi \rightarrow Pe^+e^-$, the form factors of $J/\psi \rightarrow
\eta^{\prime}\gamma^{*}$, in addition to the processes with only light flavor hadrons, such as  $P\rightarrow V\gamma$, $V\rightarrow
P\gamma$, $P\rightarrow \gamma\gamma$, $P\rightarrow \gamma
l^{+}l^{-}$, $V\rightarrow Pl^{+}l^{-}$,  as well as the form
factors of $\eta\rightarrow \gamma\gamma^{*}, \eta'\rightarrow
\gamma\gamma^{*}, \phi\rightarrow \eta\gamma^{*}$.

This paper is organized as follows. In Sec.~\ref{theor}, we
introduce the theoretical framework and elaborate the calculations
for the transition amplitudes of $J/\psi \rightarrow P\gamma^{\ast}$
and $J/\psi \rightarrow VP$. In Sec.~\ref{pheno}, we present the fit
results and discuss the interplay between different mechanisms in $J/\psi$ decays.
Summary and conclusions are given in Sec.~\ref{conclu}.

\section{Theoretical framework}
\label{theor}

\subsection{The relevant Lagrangian of $J/\psi$ hadronic decays}\label{sect.lag}

We will simultaneously study the $J/\psi$ decaying into light hadrons and the light meson
radiative decays. Therefore two different types of effective Lagrangians, i.e.
the ones involving interactions between $J/\psi$ and light hadrons, and those only including interactions between light-hadron themselves,
need to be constructed. The latter have been discussed in detail in Ref.~\cite{chen2012} and we simply introduce them below for completeness.

For the interaction operators between $J/\psi$ and light hadrons, we construct the effective Lagrangian
by taking the basic building blocks involving light hadron states from $\chi$PT~\cite{gl845} and R$\chi$T~\cite{Ecker}.
A subtlety about the description of vector resonances in R$\chi$T should be pointed out. The vector resonances are
described in the antisymmetry tensor representation~\cite{gl845,Ecker}, not in the conventional Proca field formalism.
The reason behind is that with the vector resonances in the antisymmetric tensor representation,
one can collect, upon integrating out the heavy resonance states, the bulk of low energy constants
in $\chi$PT without including the additional local counterterms~\cite{Ecker:1989yg}.
Therefore we will use the antisymmetric tensor formalism to describe the light vector resonances,
as we did in Ref.~\cite{chen2012}.  While for the $J/\psi$, we will simply
use the Proca field formalism in order to reduce the number of free couplings.

In order to set up the notations, we introduce the effective Lagrangian only involving light hadrons first.
In the large $N_C$ limit, the $U_A(1)$ anomaly from QCD is suppressed so that the singlet
$\eta_0$ meson becomes the ninth pNGB and can be systematically
incorporated into the $U(3)$ chiral Lagrangian~\cite{ua1innc,Kaiser00,Siklody}.
We use the exponential realization for $U(3)_L \times U(3)_R/ U(3)_V$ coset coordinates
\begin{eqnarray}
\tilde{U} &=&  \tilde{u}^2 = e^{i\frac{ \sqrt2\Phi}{ F}} \,,
\end{eqnarray}
where the pNGB octet plus singlet are given by
\begin{equation}
\Phi=
 \left( {\begin{array}{*{3}c}
   {\frac{1}{\sqrt{2}}\pi ^0 +\frac{1}{\sqrt{6}}\eta _8+\frac{1}{\sqrt{3}}\eta_0 } & {\pi^+ } & {K^+ }  \\
   {\pi^- } & {-\frac{1}{\sqrt{2}}\pi ^0 +\frac{1}{\sqrt{6}}\eta _8+\frac{1}{\sqrt{3}}\eta_0} & {K^0 }  \\
   { K^-} & {\overline{K}^0 } & {-\frac{2}{\sqrt{6}}\eta_8+\frac{1}{\sqrt{3}}\eta_0 }  \\
\end{array}} \right) \,.
\end{equation}
The basic building blocks involving the pNGBs and external source fields read
\begin{eqnarray}\label{defbb}
 \tilde{u}_\mu = i \tilde{u}^\dagger  D_\mu \tilde{U} \tilde{u}^\dagger \, =\,
i \{ \tilde{u}^\dagger (\partial_\mu - i r_\mu) \tilde{u}\, -\, \tilde{u}(\partial_\mu - \tilde{u}\ell_\mu) \tilde{u}^\dagger\}
\,, \nn\\
\tilde{\chi}_\pm  = \tilde{u}^\dagger   \chi \tilde{u}^\dagger \pm  \tilde{u}  \chi^\dagger  \tilde{u} \,,
 \qquad \qquad  \tilde{f}_\pm^{\mu\nu} = \tilde{u} F_L^{\mu\nu} \tilde{u}^\dagger \, \pm \,
\tilde{u}^\dagger F_R^{\mu\nu} u\, ,
\end{eqnarray}
where $\chi=2 B_0 (s + i p)$ incorporates the pseudoscalar ($p$) and scalar ($s$) external sources.
$F_L^{\mu\nu}$ and $F_R^{\mu\nu}$ are the field-strength tensors for the left and right external sources, respectively.
All of the building blocks $X=\tilde{u}_\mu, \tilde{\chi}_{\pm}, \tilde{f}^{\mu\nu}_{\pm}$ in Eq.~\eqref{defbb} then transform under the chiral group transformations as
\begin{equation}
 X \to h X h^\dagger\,, \qquad h\in U(3)_V \,.
\end{equation}
Notice that we have introduced the tildes to the objects involving the pNGB nonet in order to distinguish those with octet from $SU(3)$ $\chi$PT.
In the following construction of effective Lagrangian with light resonances and $J/\psi$,
the pNGB fields will enter only through the three types of building blocks presented in Eq.~\eqref{defbb}.

The $U(3)$ $\chi$PT Lagrangian to lowest order, $O(p^2)$, is
\begin{eqnarray}  \mathscr{L}_\chi^{(2)}=\frac{F^2}{4}\bra \tilde{u}_\mu \tilde{u}^\mu
+\tilde{\chi}_+\ket + \frac{F^2}{3}M_0^2\ln^2\det \tilde{u}\,,
\end{eqnarray}
where the last term stands for the QCD $U_A(1)$ anomaly effect, leading to a non-vanishing mass for the $\eta_0$
field even in the chiral limit.
The parameter $F$ denotes the value of the pion decay constant $F_\pi=92.2$~MeV in the
chiral limit and $B_0$ in Eq.~\eqref{defbb} is related to the quark condensate
through $\bra0\mid\psi\overline{\psi}\mid0\ket=-F^{2}B_{0}[1+O(m_{q})]$, with
$m_q$ the light quark mass. The explicit chiral symmetry breaking is
realized in $\chi$PT by assigning the vacuum expectation values of
the scalar sources as $s=$Diag$\{m_u,m_d,m_s\}$. Throughout, we take $m_u=m_d$ and use the leading order relations
$2m_u B_0=m_\pi^2$ and $(m_u+m_s)B_0=m_K^2$~\cite{Gel68,gl845}.

The physical $\eta$ and $\eta'$ states are from the mixing between $\eta_8$ and $\eta_0$.
Following a general discussion in $U(3)$ $\chi$PT, the $\eta$-$\eta^{\prime}$ mixing should be formulated in the
two-mixing-angle framework, instead of the conventional one-mixing-angle scheme~\cite{Leutwyler98,Kaiser98}.
In the octet and singlet basis, the $\eta$-$\eta^{\prime}$ mixing is parameterized by~\cite{Leutwyler98,Kaiser98}
\begin{eqnarray} \label{twoanglesmixing81}
 \left(
 \begin{array}{c}
 \eta   \\
 \eta' \\
 \end{array}
 \right) = \frac{1}{F}\left(
                                        \begin{array}{cc}
                                          F_8\, \cos{\theta_8}  & -F_0 \,\sin{\theta_0}  \\
                                           F_8\,\sin{\theta_8} & F_0 \,\cos{\theta_0} \\
                                        \end{array}
    \right)
      \left(
       \begin{array}{c}
       \eta_8   \\
       \eta_0  \\
       \end{array}
        \right)\,,
\end{eqnarray}
where $F_8$ and $F_0$ denote the weak decay constants of
the axial octet and singlet currents, respectively. By taking
$F_8=F_0=F$ and $\theta_0=\theta_8$ in Eq.(\ref{twoanglesmixing81}),
the conventional one-mixing-angle scheme is recovered.

Analogously, one can choose the quark-flavor basis to parameterize the $\eta$-$\eta'$ mixing as
\begin{eqnarray} \label{twoanglesmixingqs}
 \left(
 \begin{array}{c}
 \eta   \\
 \eta' \\
 \end{array}
 \right) = \frac{1}{F}\left(
                                        \begin{array}{cc}
                                          F_q\, \cos{\phi_q}  & -F_s \,\sin{\phi_s}  \\
                                           F_q\,\sin{\phi_q} & F_s \,\cos{\phi_s} \\
                                        \end{array}
    \right)
      \left(
       \begin{array}{c}
       \eta_q   \\
       \eta_s  \\
       \end{array}
        \right)\,,
\end{eqnarray}
with $\eta_q=(\eta_8+\sqrt2 \eta_0)/\sqrt3$ and
$\eta_s=(\eta_0-\sqrt2\eta_8)/\sqrt3$. In this case the $\eta_q$ and
$\eta_s$ states are generated by the axial vector currents with the
quark flavors $q\bar{q}=(u\bar{u}+d\bar{d})/\sqrt2$ and $s\bar{s}$,
respectively. Obviously the two mixing matrices from different bases
in Eqs.~\eqref{twoanglesmixing81} and \eqref{twoanglesmixingqs} are
related to each other through an orthogonal transformation. So they
are equivalent to describing the $\eta$-$\eta'$ mixing.  If only the
leading order of $N_C$ chiral operators with quark mass corrections
are considered, the $SU(3)$ breaking by quark masses will affect the
$\eta_q$ and $\eta_s$ differently in the quark-flavor basis, and
the angles $\phi_q$ and $\phi_s$ will be equal. This is the characteristic of the
Feldmann-Kroll-Stech (FKS) formalism~\cite{Feldmann98}.  If general
operators are included in the discussion, the quark-flavor basis
will lose these features. Nevertheless, it seems that the
phenomenological analyses support the fact that the values
of $\phi_q$ and $\phi_s$ are indeed very close to each
other~\cite{Feldmann98,Feldmann00,Escribano05}. In the following
phenomenological discussions we will explore both mixing scenarios in the singlet-octet and quark-flavor bases.

Next we follow closely R$\chi$T~\cite{Ecker} to include the vector resonances.
The ground multiplet of vector resonances was explicitly incorporated in the
antisymmetric tensor representation in R$\chi$T. The kinetic term of the vector resonance Lagrangian reads~\cite{Ecker}
\begin{eqnarray}\label{lagkinv}
\mathscr{L}_{kin}(V)=-\frac{1}{2}\bra\nabla^{\lambda}V_{\lambda\mu}\nabla_{\nu}V^{\nu\mu}-\frac{M_V^2}{2}V_{\mu\nu}V^{\mu\nu}\ket\,,
\end{eqnarray}
where the ground vector nonet matrix is given by
\begin{equation}\label{defu3v}
V_{\mu\nu}=
 \left( {\begin{array}{*{3}c}
   {\frac{1}{\sqrt{2}}\rho ^0 +\frac{1}{\sqrt{6}}\omega _8+\frac{1}{\sqrt{3}}\omega _0 } & {\rho^+ } & {K^{\ast+} }  \\
   {\rho^- } & {-\frac{1}{\sqrt{2}}\rho ^0 +\frac{1}{\sqrt{6}}\omega _8+\frac{1}{\sqrt{3}}\omega _0} & {K^{\ast0} }  \\
   { K^{\ast-}} & {\overline{K}^{\ast0} } & {-\frac{2}{\sqrt{6}}\omega_8+\frac{1}{\sqrt{3}}\omega _0 }  \\
\end{array}} \right)_{\mu\nu}\,,
\end{equation}
and the covariant derivative and the chiral connection are defined as
\begin{eqnarray}
\nabla_{\mu}V=\partial_{\mu}V+[\tilde{\Gamma}_{\mu},V],\hspace{1.5cm}
\tilde{\Gamma}_{\mu}=\frac{1}{2}\{\tilde{u}^{+}(\partial
_\mu-ir_\mu)\tilde{u}+\tilde{u}(\partial
_\mu-il_\mu)\tilde{u}^{+}\} \,.
\end{eqnarray}
The transformation laws of the resonance multiplet and its covariant derivative under chiral group transformations
are the same as the building blocks in Eq.~\eqref{defbb}
\begin{equation}
 V   \to h V h^\dagger\,, \quad  \nabla_{\mu}V \to h (\nabla_{\mu}V) h^\dagger, \qquad h\in U(3)_V \,.
\end{equation}

The masses of the resonances in the ground multiplet are degenerate in Eq.~\eqref{lagkinv} and their mass splitting
is governed by a single resonance operator at leading order of $1/N_C$~\cite{Cirigliano:2003yq}
\begin{eqnarray}
\label{lagemr}
-\,\,\,   \frac{1}{2} \, e_m^V \bra V_{\mu\nu} V^{\mu\nu} \chi_+ \ket \,.
\end{eqnarray}
It has demonstrated that the single operator in the previous equation can well explain the mass splittings of the ground
vector resonances in Eq.~\eqref{defu3v} and can also perfectly describe the quark mass dependences of the $\rho(770)$ mass
from lattice simulations~\cite{guo09prd,Guo:2014yva}. Therefore it is justified for us to simply use the physical masses for the vector resonances
in the phenomenological discussions.

The physical states of $\omega(782)$ and $\phi(1020)$ result from
the ideal mixing of $\omega_0$ and $\omega_8$
\begin{eqnarray}\label{mixv}
\omega_0=\sqrt{\frac{2}{3}}\omega-\sqrt{\frac{1}{3}}\phi\,,
\hspace{2.5cm}
\omega_8=\sqrt{\frac{2}{3}}\phi+\sqrt{\frac{1}{3}}\omega\,.
\end{eqnarray}

The transitions between the vector resonances and the photon field are described by one single operator
in the minimal version of R$\chi$T~\cite{Ecker}
\begin{eqnarray}\label{lagvpho}
\mathscr{L}_{2}(V)=\frac{F_V}{2\sqrt{2}}\bra V_{\mu\nu}\tilde{f}_{+}^{\mu\nu}\ket \,.
\end{eqnarray}

Now we construct the effective Lagrangian describing $J/\psi$
radiative decays and $J/\psi$ decaying to light hadrons. We use the Proca vector field to describe the $J/\psi$,
mainly due to the consideration of reducing the number of coupling vertices.
We first consider the strong interaction vertices for $J/\psi$ decaying to a light vector and a pNGB. Three terms are introduced
\begin{eqnarray}
\mathscr{L}_{\psi VP}=M_\psi h_1 \varepsilon_{\mu\nu\rho\sigma} \psi^\mu\bra \tilde{u}^\nu V^{\rho\sigma}\ket
+\frac{1}{M_\psi} h_2 \varepsilon_{\mu\nu\rho\sigma}
\psi^\mu\bra \{\tilde{u}^\nu, V^{\rho\sigma}\}\tilde{\chi}_+\ket  +M_\psi
h_3\varepsilon_{\mu\nu\rho\sigma}\psi^\mu\bra \tilde{u}^\nu\ket \bra V^{\rho\sigma}\ket \,,\label{lagpsivp}
\end{eqnarray}
where the first term can be related to the leading three-gluon-annihilation (singly OZI disconnected) diagram  proposed
in Refs.~\cite{haber85,seiden88,Ball,bramon97,QZhao2007,QZhao2008,Thomas,Escribano2010}, the second term
stands for the strong $SU(3)$ symmetry breaking term caused by the quark masses and the last one corresponds to
the doubly OZI-suppressed diagram. We have introduced the $M_{\psi}$ factors in Eq.~\eqref{lagpsivp} so that
the couplings $h_{i=1,2,3}$ are dimensionless.

For the interaction vertices with $J/\psi$, one pNGB and one photon field, we have two operators
\begin{eqnarray}
\mathscr{L}_{\psi P\gamma }=g_1 \varepsilon_{\mu\nu\rho\sigma} \psi^\mu\bra \tilde{u}^\nu \tilde{f}_+^{\rho\sigma}\ket
+\frac{1}{M^2_\psi} g_2 \varepsilon_{\mu\nu\rho\sigma}
\psi^\mu\bra \{\tilde{u}^\nu,
\tilde{f}_+^{\rho\sigma}\}\tilde{\chi}_+\ket ,\label{lagpsigp}\end{eqnarray}
where the second term generates the $SU(3)$-flavor symmetry breaking
caused by the quark masses for the $J/\psi\, P\gamma$ vertices, with $P=\pi, \eta, \eta'$.

The transition between the $J/\psi$ and the photon field is described by
\begin{eqnarray} \label{lagpsipho}
\mathscr{L}_2^\psi=\frac{-1}{2\sqrt{2}}\frac{f_\psi}{M_\psi}\bra \hat{\psi}_{\mu\nu} \tilde{f}_+^{\mu\nu}\ket \,,
\end{eqnarray}
with $\hat{\psi}_{\mu\nu}=\partial_\mu \psi^\nu-\partial_\nu
\psi^\mu$. The coupling strength $f_\psi$ can be determined from the
decay width of $J/\psi\rightarrow e^+ e^-$:
\begin{eqnarray} f_\psi=\left(\frac{27 M_\psi \Gamma_{\psi\rightarrow e^+ e^-}}{32\pi \alpha^2}\right)^{\frac{1}{2}},\end{eqnarray}
where the masses of electron and positron have been neglected, and
$\alpha =e^2/4\pi$ stands for the fine structure constant.

\begin{figure}[ht]
\centering
\includegraphics[height=4cm,width=15cm]{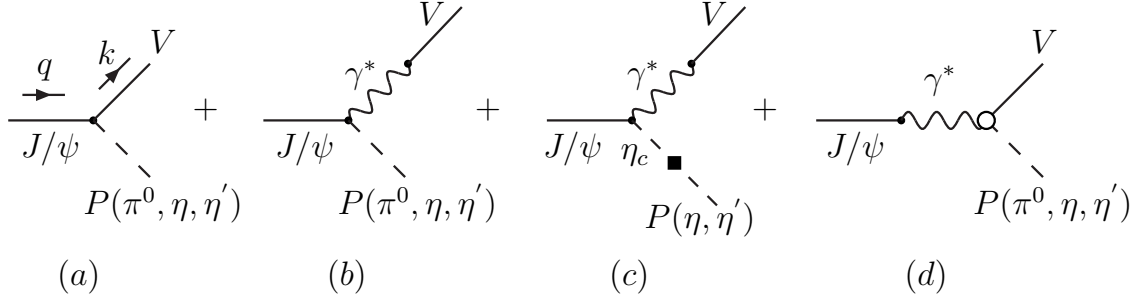}
\caption{ Feynman diagrams for the processes $J/\psi \rightarrow
VP$. The notations of the solid square in diagram (c) and the open circle in diagram (d) are explained in the text.
}\label{fig.psivp}
\end{figure}

\begin{figure}[ht]
\centering
\includegraphics[height=4cm,width=15cm]{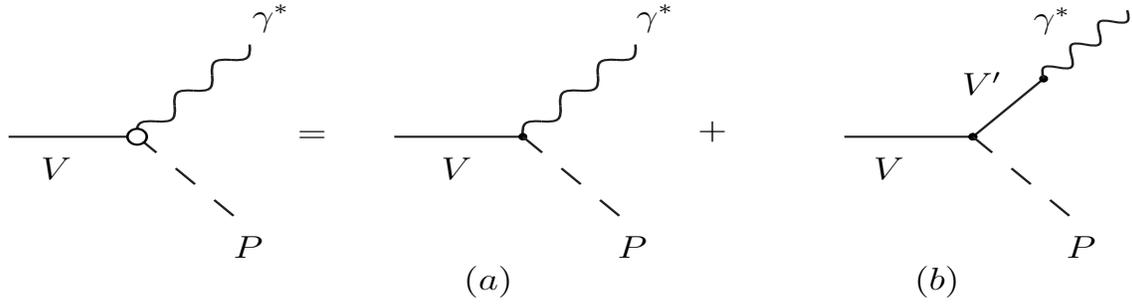}
\caption{ Diagrams relevant to the $V\rightarrow P \gamma^{*}$
processes: (a) direct type and (b) indirect type. }\label{fig.vpg}
\end{figure}

The $J/\psi\rightarrow VP$ decay processes can be categorized into
two classes: (i) isospin conserved channels, such as
$J/\psi\rightarrow \rho\pi, \omega\eta^{(\prime)},
\phi\eta^{(\prime)}, K^{\ast} \bar{K}$, which include both strong
and EM transitions; (ii) isospin violated channels, such as
$J/\psi\rightarrow \rho\eta^{(\prime)},\omega\pi^{0},$ of which the
leading contribution is the EM transition.  In Fig.~\ref{fig.psivp},
we show the four types of diagrams that contribute to the
$J/\psi\rightarrow VP$ decays, where the diagram (a) represents the
strong interactions from the Lagrangian in Eq.~\eqref{lagpsivp}, and
the remaining diagrams (b-d) depict the EM interactions. The solid square denotes
the mixing between $\eta_c$ and $\eta(\eta')$, which will be addressed in the following section. The open
circle in the diagram (d) of Fig.~\ref{fig.psivp} stands for the
radiative transition amplitudes of the pNGBs and light vector
resonances, which have been the focus in our previous work in
Ref.~\cite{chen2012}. We will directly take these amplitudes from
the former reference.  For the sake of completeness, we simply show
the effective Lagrangians that are relevant to the radiative
transition amplitudes of the pNGBs and light vector resonances
below. We refer to Ref.~\cite{chen2012} for details about the
constructions of these Lagrangians. All of the Lagrangians are
constructed in the framework of R$\chi$T.  In our scheme the
$VP\gamma^{*}$ transition receives two types of contributions: the
contact diagram and the resonance-exchange one, as shown in
Fig.~\ref{fig.vpg}. 
The chiral effective Lagrangians with antisymmetric tensor formalism for the light 
vector resonances that are pertinent to these kinds of processes, are first written down in Ref.~\cite{Femenia} and 
then completed in a more general setting in Ref.~\cite{kampf11prd}. The focus of the previous two references is the 
$SU(3)$ case with the light pseudoscalar octet. We generalize the relevant discussions 
to the $U(3)$ case in Ref.~\cite{chen2012}, so that we can study the processes involving $\eta$ and $\eta'$ states.   
The $U(3)$ operators with one vector resonance, one external source and one pNGB are given by 
\begin{eqnarray}\label{LagVJP}
\mL_{VJP}=&\frac{\tilde{c}_1}{M_V}&\varepsilon_{\mu\nu\rho\sigma}
\langle \{V^{\mu\nu},\tilde{f}_+^{\rho\alpha}\}\nabla_\alpha
\tilde{u}^\sigma \rangle
+\frac{\tilde{c}_2}{M_V}\varepsilon_{\mu\nu\rho\sigma}\langle
\{V^{\mu\alpha},\tilde{f}_+^{\rho\sigma}\}\nabla_\alpha
\tilde{u}^\nu\rangle
+\frac{i\tilde{c}_3}{M_V}\varepsilon_{\mu\nu\rho\sigma}
\langle\{V^{\mu\nu},\tilde{f}_+^{\rho\sigma}\} \tilde{\chi}_-\rangle \nonumber\\
&+&\frac{i\tilde{c}_4}{M_V}\varepsilon_{\mu\nu\rho\sigma}\langle
V^{\mu\nu}[\tilde{f}_-^{\rho\sigma},\tilde{\chi}_+]\rangle
+\frac{\tilde{c}_5}{M_V}\varepsilon_{\mu\nu\rho\sigma}\langle
\{\nabla_\alpha
V^{\mu\nu},\tilde{f}_+^{\rho\alpha}\}\tilde{u}^\sigma\rangle
+\frac{\tilde{c}_6}{M_V}\varepsilon_{\mu\nu\rho\sigma}\langle
\{\nabla_\alpha
V^{\mu\alpha},\tilde{f}_+^{\rho\sigma}\}\tilde{u}^\nu\rangle \nonumber\\
&+&\frac{\tilde{c}_7}{M_V}\varepsilon_{\mu\nu\rho\sigma}\langle
\{\nabla^\sigma
V^{\mu\nu},\tilde{f}_+^{\rho\alpha}\}\tilde{u}_\alpha\rangle
-i\tilde{c}_8M_V
\sqrt{\frac{2}{3}}\varepsilon_{\mu\nu\rho\sigma}\langle
V^{\mu\nu}\tilde{f}_+^{\rho\sigma}\rangle \ln(\det\tilde{u})\,,
\end{eqnarray}
which are responsible for the contact diagram in Fig.~\ref{fig.vpg}.
For the resonance-exchange diagram, the responsible effective Lagrangian reads
\begin{eqnarray}\label{LagVVP}
\mL_{VVP}=&\tilde{d}_{1}&\varepsilon_{\mu\nu\rho\sigma}\langle
\{V^{\mu\nu},V^{\rho\alpha}\}\nabla_\alpha \tilde{u}^\sigma\rangle +
i\tilde{d}_{2}\varepsilon_{\mu\nu\rho\sigma}\langle
\{V^{\mu\nu},V^{\rho\sigma}\}\tilde{\chi}_-\rangle
+\tilde{d}_{3}\varepsilon_{\mu\nu\rho\sigma}\langle \{\nabla_\alpha
V^{\mu\nu},V^{\rho\alpha}\}\tilde{u}^\sigma\rangle \nonumber\\
&+&\tilde{d}_{4}\varepsilon_{\mu\nu\rho\sigma}\langle
\{\nabla^\sigma V^{\mu\nu},V^{\rho\alpha}\}\tilde{u}_\alpha\rangle
-i\tilde{d}_5M_V^2\sqrt{\frac{2}{3}}\varepsilon_{\mu\nu\rho\sigma}\langle
 V^{\mu\nu}V^{\rho\sigma}\rangle \ln(\det\tilde{u})\,.
\end{eqnarray}

To impose the high energy constraints to the couplings can greatly reduce the
number of free parameters in R$\chi$T~\cite{Ecker:1989yg,Roig:2014uja,Guo:2010dv}.
This procedure also renders the amplitudes calculated in R$\chi$T consistent with the behavior dictated by QCD.
We will follow this procedure in this work as well. Through matching the leading operator product expansion (OPE) of the
$VVP$ Green function with the result evaluated within R$\chi$T and
requiring the vector form factor to vanish in the high energy limit, we
obtain the high energy constraints on resonance couplings~\cite{chen2012}
\begin{eqnarray}  
4\tilde{c}_3+\tilde{c}_1&=&0, \nonumber\\
\tilde{c}_1-\tilde{c}_2+\tilde{c}_5&=&0, \nonumber\\
\tilde{c}_5-\tilde{c}_6&=&\frac{N_C}{64\pi^2}\frac{M_V}{\sqrt{2}F_V}, \nonumber\\
\tilde{d}_1+8\tilde{d}_2 -\tilde{d}_3 &=& \frac{F^2}{8F_V^2}, \nonumber\\
\tilde{d}_3&=&-\frac{N_C}{64\pi^2}\frac{M_V^2}{F_V^2}\nonumber\\
\tilde{c}_8 &=&  -\frac{ \sqrt2 M_0^2}{\sqrt3 M_V^2}
\tilde{c}_1 \,. \label{he-ope-c8}
\end{eqnarray}

\subsection{The transition amplitudes of $J/\psi \rightarrow P\gamma^{\ast}$}

\begin{figure}[h]
\includegraphics[height=4cm,width=15cm]{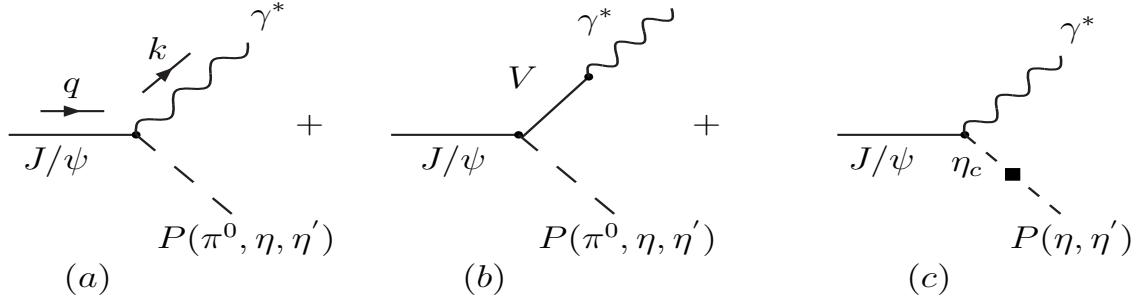}
\caption{ Feynman diagrams for the processes $J/\psi \rightarrow P
\gamma^*$ }\label{fig.psipg}
\end{figure}

In accord with the covariant Lorentz structure, the general amplitude for the
radiative decay $J/\psi(q) \to P(q-k)\gamma^\ast(k)$ takes the form
\begin{eqnarray}\label{defgpg}
i\mathcal{M}_{\psi\to P\gamma^\ast}= i e\,
\varepsilon_{\mu\nu\rho\sigma} \epsilon_\psi^\mu
\epsilon_{\gamma^\ast}^\nu q^\rho k^\sigma
 G_{\psi \to P\gamma^\ast}(s)\,,
\end{eqnarray}
where $q$ and $k$ stand for the four momenta of $J/\psi$ and
$\gamma^{\ast}$ respectively; $s=k^2$; $\epsilon_{\psi}$ and
$\epsilon_{\gamma^{\ast}}$ are the polarization vectors; $e$ is the electric charge of a positron.
The relevant Feynman diagrams to the radiative decay $J/\psi \to P\gamma^\ast$ are displayed in Fig.~\ref{fig.psipg}.
Using the previously introduced Lagrangian in Sec.~\ref{sect.lag},
it is straightforward to calculate the contributions from the diagrams
(a) and (b) in Fig.~\ref{fig.psipg} to $G_{\psi \to P\gamma^\ast}(s)$.
While for the diagram (c), we need to provide extra terms.

Based on the QCD axial anomaly and the PCAC hypothesis,
the mixing angle of the $\eta(\eta^{\prime})-\eta_c$ was evaluated in Ref.~\cite{KTChao1990} and it was found that the mechanism from the
diagram (c) in Fig.~\ref{fig.psipg} dominates the $J/\psi \rightarrow \eta(\eta^{\prime}) \gamma$ decays.
The contribution from this diagram to the $J/\psi \rightarrow \eta(\eta^{\prime}) \gamma$ amplitude can be generally written as
\begin{eqnarray}\label{etacmixing}
i\mathcal{M}_{\psi\to \eta(\eta^{\prime})\gamma^{\ast}}^{mixing}= i
e\, \varepsilon_{\mu\nu\rho\sigma} \epsilon_\psi^\mu
\epsilon_{\gamma^{\ast}}^\nu q^\rho k^\sigma \,
 \lambda_{P\eta_c} \, g_{\psi \eta_c\gamma}(s) \, e^{i\delta_P}\,,
\end{eqnarray}
where the mixing strengths are obtained as $\lambda_{\eta\eta_c}=-4.6\times
10^{-3}, \lambda_{\eta^{\prime}\eta_c}=-1.2\times
10^{-2}$ in Ref.~\cite{KTChao1990}. $\delta_P$, with $P=\eta, \eta'$, stand for the relative phases between diagram (c) and others, which
are free parameters in this work and will be fitted later. The coupling strength $g_{\psi \eta_c\gamma}(s)$ is defined as
\begin{eqnarray}
i\mathcal{M}_{\psi\to \eta_c\gamma^*}= i e\,
\varepsilon_{\mu\nu\rho\sigma} \epsilon_\psi^\mu
\epsilon_{\gamma^*}^\nu q^\rho k^\sigma
 g_{\psi \eta_c\gamma}(s)\,,
\end{eqnarray}
and we can easily obtain
\begin{eqnarray}
|g_{\psi \eta_c\gamma}(0)| =\left(\frac{24 M_\psi^3 \Gamma_{\psi\rightarrow \eta_c
\gamma}}{\alpha(M_\psi^2-m_{\eta_c}^2)^3}\right)^{\frac{1}{2}}.
\end{eqnarray}
Since we focus on the $J/\psi \to \eta_c \gamma^* \to \eta_c l^+ l^-$ process, the interval of the energy squared $s$
is limited to a small region, comparing with the scale $M_\psi^2$.
This justifies us to use the spacelike form factor derived in Ref.~\cite{Dudek2006}
\begin{eqnarray}
\label{Eqpsietacg}  g_{\psi \eta_c\gamma}(s)
=g_{\psi \eta_c\gamma}(0)e^{\frac{s}{16\beta^2}} \,,
\end{eqnarray}
to the $s\sim 0$ timelike region, due to the continuity condition of the form factor at $s=0$.
In order not to interrupt the present discussions, the lengthy expressions
of $G_{\psi \to P\gamma^\ast}(s)$ calculated from Fig.~\ref{fig.psipg}
are relegated in Appendix~\ref{ffpsipg}.

In our convention, the decay widths of $J/\psi \to P\gamma$ are
\begin{eqnarray}
\Gamma(\psi\rightarrow
P\gamma)=\frac{1}{3}\alpha\left(\frac{M_\psi^2-M_P^2}{2M_\psi}\right)^3|G_{\psi\rightarrow
P\gamma^{*}}(0)|^2\,,
\end{eqnarray}
and the decay widths of $J/\psi \to P\gamma^\ast \to P l^+ l^-$ are
\begin{eqnarray}\label{gammapsipll}
\Gamma_{ \psi \to P l^+ l^-}= \int^{(M_\psi-m_P)^2}_{4m_l^2}
\frac{\alpha^2(2 m_l^2+s)}{72 M_\psi^3 \pi
s^3}\sqrt{s(s-4m_l^2)} \, \big[\lambda(s,M_\psi,m_P)]^3 \,
|G_{\psi \to P\gamma^\ast}(s)|^2 d s,
\end{eqnarray}
where the leptons $l=e, \mu$, the pNGBs $P=\eta, \eta'$ and
\begin{equation}
\lambda(s,M_\psi,m_P)= \sqrt{[s-(M_\psi-m_P)^2][s-(M_\psi+m_P)^2]}\,.
\end{equation}

\subsection{Transition amplitudes of the $J/\psi \rightarrow VP$ processes}

The general amplitude for the decay $J/\psi(q) \to V(k) P(q-k)$ in accord with the Lorentz structure can
be written as
\begin{eqnarray}\label{defgvp}
i\mathcal{M}_{\psi\to VP}= i \, \varepsilon_{\mu\nu\rho\sigma}
\epsilon_\psi^\mu \epsilon_V^\nu q^\rho k^\sigma
 G_{\psi \to VP}\,,
\end{eqnarray}
and the relevant Feynman diagrams are displayed in Fig.~\ref{fig.psivp}.
The diagram (a) in Fig.~\ref{fig.psivp} represents the strong interaction part,
which can be evaluated with the Lagrangian in Eq.~\eqref{lagpsivp}.
The remaining diagrams receive EM contributions,
which can be evaluated with the Lagrangians in Eqs.~\eqref{lagvpho}\eqref{lagpsigp}\eqref{lagpsipho}\eqref{LagVJP}\eqref{LagVVP}
and the amplitude in Eq.~\eqref{etacmixing}.

For the diagram (d) in Fig.~\ref{fig.psivp}, we need the amplitudes of $V(k)P(q-k)\gamma^{*}(q)$, which are depicted in Fig.~\ref{fig.vpg}
 and can be written as
\begin{eqnarray}\label{defFF}
i\mathcal{M}_{VP\gamma^{*}}= -i \,e\, \varepsilon_{\mu\nu\rho\sigma}
\epsilon_V^\mu \epsilon_{\gamma^{*}}^\nu q^\rho k^\sigma
F_{VP\gamma^{*}}(s=q^2)\,.
\end{eqnarray}
The form factors $F_{VP\gamma^{*}}(s)$ are calculated within the framework of
R$\chi$T in Ref.~\cite{chen2012} and we simply give the results in Appendix~\ref{ffvpg} for completeness.
In this convention, the contribution from the diagram (d) in Fig.~\ref{fig.psivp} to the $G_{\psi \to VP}$ is found to be
$\frac{8\sqrt{2}\pi\alpha}{3}\frac{ f_\psi}{M_\psi}F_{VP\gamma^{*}}(s=M_\psi^2)$.

The full expressions of $G_{\psi \to VP}$ from Fig.~\ref{fig.psivp} are relegated in Appendix~\ref{ffpsivp}.
The decay widths of $J/\psi \to VP$ are given by
\begin{eqnarray}
\Gamma(\psi\rightarrow VP)=\frac{1}{96\pi
M_\psi^3}\bigg\{ \big[M_\psi^2-(M_V-m_P)^2\big]\big[M_\psi^2-(M_V+m_P)^2\big]\bigg\}^{\frac{3}{2}} \,\,\big|G_{\psi\rightarrow VP}\big|^2\,.
\end{eqnarray}

\section{ Phenomenology discussion } \label{pheno}

The experimental data that we consider in this work include the
decay widths of $J/\psi \to P\gamma$ and $J/\psi \to VP$~\cite{Pdg},
with $P=\pi, K, \eta, \eta'$ and $V=\rho,K^*,\omega,\phi$, and the
recently measured Dalitz decay widths of $J/\psi \to Pe^+ e^-$ and
the form factor $|F_{\psi\eta^{'}}(s)|^2=\left|\frac{G_{\psi \to
\eta^{'}\gamma^\ast}(s)}{G_{\psi \to
\eta^{'}\gamma^\ast}(0)}\right|^2$~\cite{BESIII2014}. \footnote{In
Ref.~\cite{BESIII2014}, the peaking backgrounds arising from
$J/\psi\to PV \to Pe^+e^- $, with $V=\rho^0,\omega,\phi$, are
subtracted. We acknowledge Xin-Kun Chu for patient explanations on
this issue. In accord with the experimental measurements, we do not
consider the contribution from the diagram (b) in
Fig.~\ref{fig.psipg} when fitting the data from
Ref.~\cite{BESIII2014}. To be consistent, the theoretical
predictions for the decay widths of $J/\psi \to P\mu^+\mu^-$ given
in Table~\ref{table-JpsitoPll} are obtained by excluding the diagram
(b) in Fig.~\ref{fig.psipg}.}  Also we take into account all the
radiative decay processes considered in our previous paper with only
light hadron states~\cite{chen2012}: $P\rightarrow V\gamma$,
$V\rightarrow P\gamma$, $P\rightarrow \gamma\gamma$, $P\rightarrow
\gamma l^{+}l^{-}$, $V\rightarrow Pl^{+}l^{-}$, as well as the form
factors of $\eta\rightarrow \gamma\gamma^{*}, \eta'\rightarrow
\gamma\gamma^{*}, \phi\rightarrow \eta\gamma^{*}$. Below we will
make a global fit by taking the two types of data together, i.e.
these with the $J/\psi$ and those without the $J/\psi$.

\begin{figure}[ht]
\includegraphics[height=8cm,width=12cm]{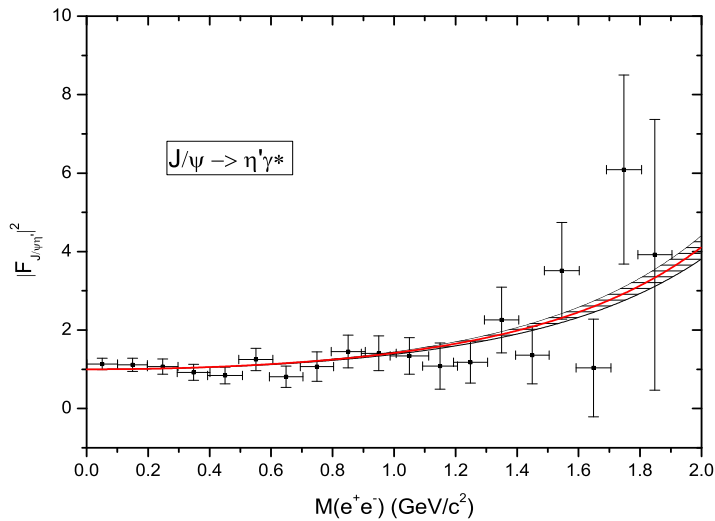}
\caption{ (Color online). The form factors of $ J/\psi \rightarrow
\eta^{\prime}\gamma^{*}$. The red solid line corresponds to the result
with the central values of the parameters in Table~\ref{tablepar} and the shaded areas stand for the error bands.
The experimental data are taken from~\cite{BESIII2014}.
\label{fig.Jpsietapff} }
\end{figure}

For the resonance operators $c_i$ and $d_j$ in Eqs.~\eqref{LagVJP} and \eqref{LagVVP},  
we impose the high energy constraints presented in Eq.~\eqref{he-ope-c8}, in such a way it reduces 
six combinations of unknown parameters and is quite helpful to stabilize the fit. In addition, it makes the asymptotic 
behaviors of the relevant amplitudes consistent with QCD in the large $N_C$ and chiral limits. 
The $\tilde{c}_4$ term in Eq.~\eqref{LagVJP}, which contributes
exclusively to the vertex $K^{\ast\pm} K^{\pm}\gamma$, is the focus of Ref.~\cite{chen2014},
and $\tilde{c}_4=-0.0023$ is determined there. In Ref.~\cite{chen2012}, a very strong linear
correlation between $\tilde{d}_2$ and $\tilde{d}_5$ is observed:
$\tilde{d}_5=4.4\tilde{d}_2-0.06$. We will take this value of
$\tilde{c}_4$ and use the linear correlation between $\tilde{d}_2$
and $\tilde{d}_5$ in our current study, and we point out that if the
values of $\tilde{c}_4, \tilde{d}_2 $ and $\tilde{d}_5$ are fitted,
the results turn out to be very close to these constraints.
For $\beta$ in Eq.~\eqref{Eqpsietacg}, we will take the value $\beta=580\pm19$
MeV~\cite{CLiu2011}. Our fitting quality is not sensitive to the
value of $\beta$ if $\beta$ is above $500$ MeV. The
reason behind is that for the $J/\psi \rightarrow \eta(\eta^{'}) l^+l^-$ decays the dominant contributions come from the region with small
values of $s$, due to the kinematic factors in Eq.~\eqref{gammapsipll}. In addition, the experiment data for the $J/\psi \rightarrow \eta^{'}
\gamma^{\ast}$ form factors have large errors, as shown in Fig.~\ref{fig.Jpsietapff}.

Before stepping into the detail of the fits, we point out that with the theoretical setups in Sect.~\ref{theor}
there are always large discrepancies between our theoretical output and the experimental measurement
for the isospin-violated channel $J/\psi \rightarrow \omega\pi^0$.  Similarly large discrepancies have also been found
between the conventional VMD approach and the experimental data of $\omega\rightarrow
\pi^0\gamma^*$ form factors~\cite{NA6009,NA6011}, and one possible solution is to include more excited resonances~\cite{Kubis,chen2014}.
Therefore in order to reasonably describe the $J/\psi \rightarrow \omega\pi^0$ decay, we simply introduce another excited vector
resonance $\rho'$ in this channel. To be more specific, we introduce the term $r_1 M_\omega D_{\rho^{'}}(s)$ to the form factor
$F_{\omega\pi\gamma^\ast}(s)$ in Eq.~\eqref{defFF} and the definition of the propagator $D_{\rho^{'}}(s)$ is given in Eq.~\eqref{defdr}.
The parameter $r_1$ will be fitted. The mass and width of $\rho^{'}$ will be fixed at $M_{\rho^{'}}=1600$ MeV and $\Gamma_{\rho^{'}}=500$ MeV, respectively.
We point out that a sizable variation of $M_{\rho^{'}}$ and $\Gamma_{\rho^{'}}$ barely affects the fitting results.

\begin{figure}[ht]
\includegraphics[height=8cm,width=12cm]{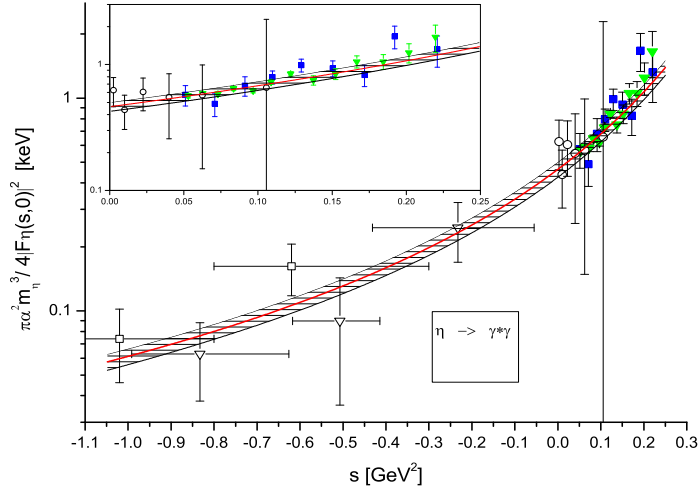}
\caption{ (Color online). The form factors of $ \eta\rightarrow \gamma^{*}\gamma$.
The red solid line corresponds to the result
with the central values of the parameters in Table~\ref{tablepar} and the shaded areas stand for the error bands.
The references of different experimental data are: solid squares~\cite{LG1,LG2}, open squares
~\cite{CELLO}, open circles ~\cite{SND}, solid triangles~\cite{NA60} and open triangles~\cite{TPC}.
The framed figure is the close-up of the plot in the region of $s>0$.  \label{fig.etaff} }
\end{figure}

\begin{figure}[h]
\includegraphics[height=8cm,width=12cm]{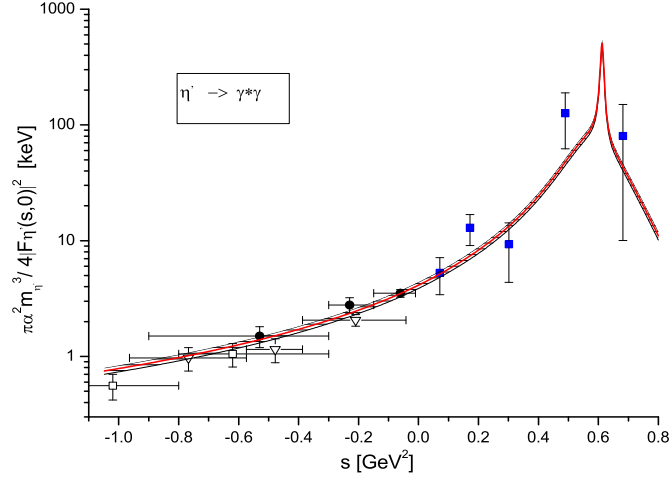}
\caption{ The form factors of $ \eta'\rightarrow \gamma^{*}\gamma$.
The red solid line corresponds to the result
with the central values of the parameters in Table~\ref{tablepar} and the shaded areas stand for the error bands.
The references of different experimental data are: solid squares~\cite{LG1,LG2}, open squares
~\cite{CELLO}, open triangles~\cite{TPC}, solid circles ~\cite{L3}.
\label{fig.etapff}}
\end{figure}

\begin{figure}[h]
\includegraphics[height=8cm,width=12cm]{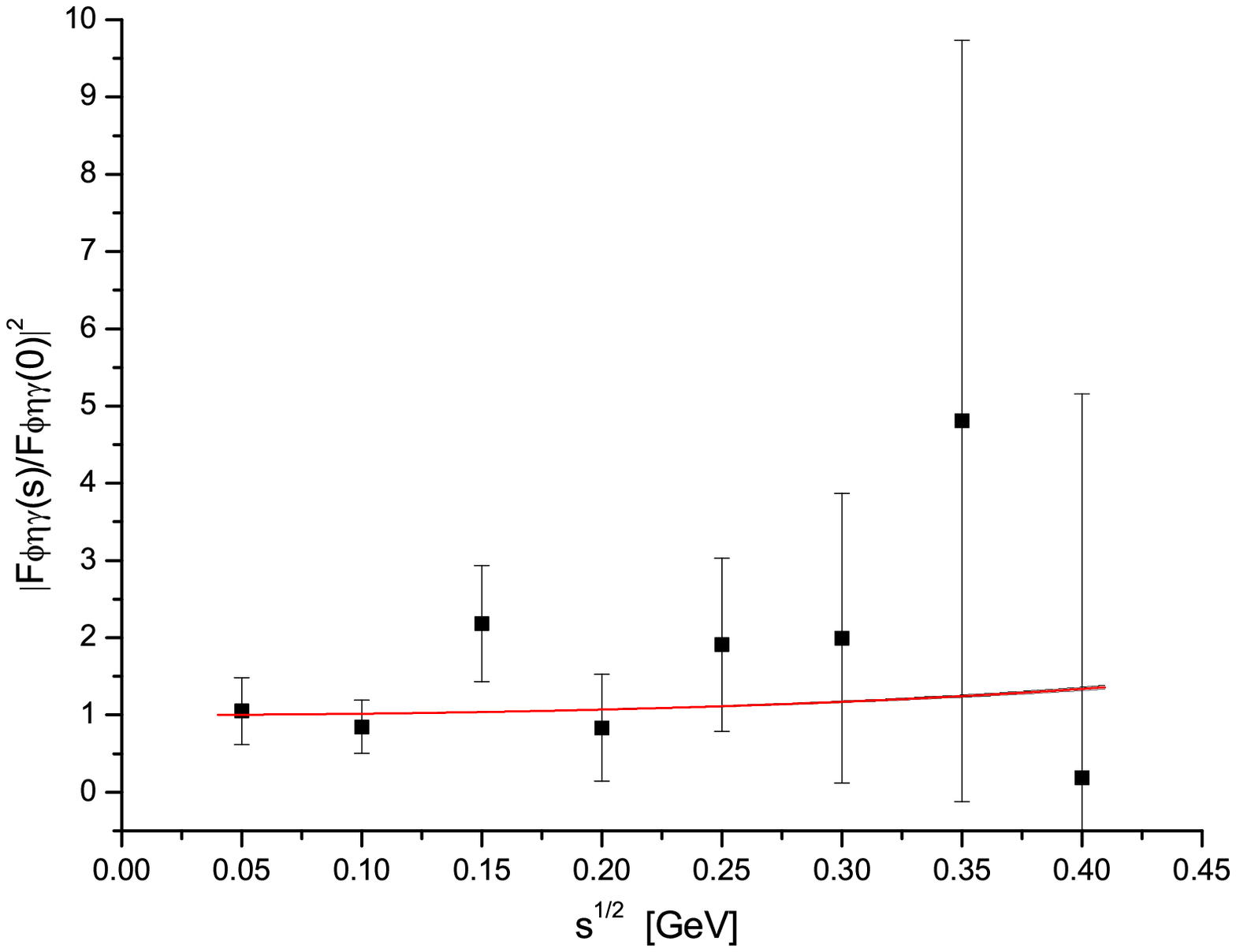}
\caption{ The form factors of $\phi\rightarrow
\eta\gamma^{*}$~\cite{SND}. The red solid line corresponds to the result
with the central values of the parameters in Table~\ref{tablepar} and the shaded areas stand for the error bands.
\label{fig.phietaff} }
\end{figure}

With all of the previous setups, it is ready to present our fit
results. The final values for the fitted parameters are given in
Table~\ref{tablepar}. For the various decay widths, we summarize the
experiment data and the results from our theoretical outputs in
Table~\ref{table-Jpsidecay} for the $J/\psi$ decay processes and in
Table~\ref{table-NoJpsidecay} for the light-hadron decays. The
resulting plots for the form factors of $J/\psi\eta'\gamma^*$,
$\eta\gamma\gamma^*$, $\eta'\gamma\gamma^*$ and $\phi\eta\gamma^*$
together with the corresponding experimental data are given in
Figs.~\ref{fig.Jpsietapff},~\ref{fig.etaff},~\ref{fig.etapff}
and~\ref{fig.phietaff}, respectively. The error bands shown in the
plots and the errors of the physical quantities in the following tables
correspond to the statistical uncertainties at one standard
deviation~\cite{etkin82prd}: $n_\sigma = (\chi^2 -
\chi_0^2)/\sqrt{2\chi_0^2}$, with $\chi_0^2$ the minimum $\chi^2$
obtained in the fit and $n_\sigma$ the number of standard
deviations.

\begin{table}[ht]
\begin{scriptsize}
\begin{center}
\begin{tabular}{|c||c|c|}
\hline
         & Fit  &  Fit in Ref.~\cite{chen2012} \\
\hline \hline
$F_8 $     &   $ (1.45\pm 0.04)F_\pi $&   $ (1.37\pm 0.07)F_\pi $   \\
$F_0$     &   $ (1.28\pm 0.06)F_\pi $  &   $ (1.19\pm 0.18)F_\pi $  \\
$\theta_8$   &       $ (-26.7\pm 1.8 )^{\circ} $ &       $ (-21.1\pm 6.0 )^{\circ} $    \\
$\theta_0$     &      $ (-11.0\pm 1.0 )^{\circ} $ &       $ (-2.5\pm 8.2 )^{\circ} $   \\
$F_V$     &      $ 134.9\pm 3.2 $  &      $ 136.6\pm 3.5 $  \\
$\tilde{c}_3$       &        $0.0029\pm 0.0006$ &        $0.0109\pm 0.0161$ \\
$\tilde{d}_2$         &        $0.081\pm 0.006$ &        $0.086\pm 0.085$ \\
$h_1$         &     $ (-2.36\pm 0.13 )\times 10^{-5}  $  & -  \\
$h_2$         &     $ (-4.73\pm 1.26 )\times 10^{-5}  $   & - \\
$h_3$         &        $(3.85\pm 0.45 )\times 10^{-6}   $  & - \\
$g_1$         &     $ (-2.92\pm 0.17 )\times 10^{-5}  $   & - \\
$g_2$         &     $ (5.93\pm 1.04 )\times 10^{-4}  $  & -  \\
$r_1$         &     $ 0.44\pm 0.10  $  & -  \\
$\delta_\eta$         &        $(39\pm 44 )^{\circ}   $  & - \\
$\delta_{\eta^{'}}$         &     $(115\pm 13 )^{\circ}  $  & -\\
&  &\\
$\frac{\chi^2}{d.o.f}$ &  $\frac{96.0}{106-15}=1.06$   &
$\frac{64.0}{70-8}=1.03$\\
\hline
$F_q $     &   $ (1.15\pm 0.04)F_\pi $&   -   \\
$F_s$     &   $ (1.56\pm 0.06)F_\pi $  &   -  \\
$\phi_q$   &       $ (34.5\pm 1.8 )^{\circ} $ &       -    \\
$\phi_s$     &      $ (36.0\pm 1.4 )^{\circ} $ &       -   \\
\hline
\end{tabular}
\caption{\label{tablepar} The parameters result from the fit. For
comparison, we provide the  results of Ref.~\cite{chen2012} which are obtained by
only fitting the light hadrons radiative decay processes. }
\end{center}
\end{scriptsize}
\end{table}

\begin{table}[h]
\begin{small}
\begin{center}
\begin{tabular}{cccc}\hline\hline
&Exp &           Fit    \\
\hline
$\psi \to \rho^0\pi^0$ &  $5.3\pm0.7$ & $5.6\pm 0.7$ \\
$\psi \to \rho\pi$ &  $16.9\pm1.5$  & $16.4\pm 1.9$  \\
$\psi \to \rho^0\eta$ &   $0.193\pm0.023$  & $0.202\pm 0.047$ \\
$\psi \to \rho^0\eta^{'}$ &   $0.105\pm0.018$  & $0.110\pm0.035$  \\
$\psi \to \omega\pi^0$ & $0.45\pm0.05$  & $0.45\pm0.12$ \\
$\psi \to \omega\eta$ & $1.74\pm 0.20$  &  $1.74\pm 0.25$  \\
$\psi \to \omega\eta^{'}$ & $0.182\pm 0.021$     & $0.184\pm 0.040$ \\
$\psi \to \phi\eta$ & $0.75\pm 0.08$   &  $0.82\pm 0.11$  \\
$\psi \to \phi\eta^{'}$ & $0.40\pm 0.07$   &  $0.38\pm 0.13$  \\
$\psi \to K^{\ast +} K^- $ + c.c. & $5.12\pm 0.30 $  &  $4.79\pm$ 0.51 \\
$\psi \to K^{\ast 0} \bar{K}^0 $ + c.c. & $4.39\pm 0.31 $  &  $4.43\pm 0.38$  \\
$\psi \to \pi^0\gamma$ & $0.0349\pm 0.0032$  &  $0.0303\pm 0.0086$  \\
$\psi \to \eta\gamma$ & $1.104\pm 0.034$   &  $1.101\pm 0.079$  \\
$\psi \to \eta^{'}\gamma$ & $5.16\pm 0.15$   &  $5.22\pm 0.15$  \\
$\psi \to \pi^0 e^+ e^-$ & $(0.0756\pm 0.0141)\times 10^{-2}$    &  $(0.1191\pm 0.0138)\times 10^{-2}$  \\
$\psi \to \eta e^+ e^-$ & $(1.16\pm 0.09)\times 10^{-2}$   &  $(1.16\pm 0.08)\times 10^{-2}$  \\
$\psi \to \eta^{'}e^+ e^-$ & $(5.81\pm 0.35)\times 10^{-2}$  &  $(5.76\pm 0.16)\times 10^{-2}$  \\
\hline\hline
\end{tabular}
\caption{\label{table-Jpsidecay}  Experimental and theoretical
values of the branching fractions ($\times
10^{-3}$) of various processes: $J/\psi \to
VP$, $J/\psi \to P\gamma$, and $J/\psi \to Pe^+ e^-$. The experimental data are taken
from~\cite{Pdg,BESIII2014}. The error bands of the theoretical outputs are
calculated by using the parameter configurations in Table~\ref{tablepar}. }
\end{center}
\end{small}
\end{table}

\begin{table}[h]
\begin{small}
\begin{center}
\begin{tabular}{cccc}\hline\hline
&Exp &           Fit    \\
\hline
$\Gamma_{\omega\rightarrow\pi\gamma}$ &  $757\pm28$ & $750\pm33$ \\
$\Gamma_{\rho^0\rightarrow\pi^0\gamma}$ &  $89.6\pm12.6$  & $78.0\pm3.4$  \\
$\Gamma_{K^{*0}\rightarrow K^0\gamma}$ &   $116\pm12$  & $116\pm5$ \\
$\Gamma_{\omega\rightarrow\eta\gamma}$ & $3.91\pm 0.38$  &  $5.16\pm 0.41$  \\
$\Gamma_{\rho^0\rightarrow\eta\gamma}$ & $44.8\pm 3.5$     & $42.6\pm 3.5$ \\
$\Gamma_{\phi\rightarrow \eta\gamma}$ & $55.6\pm 1.6$   &  $55.4\pm 3.7$  \\
$\Gamma_{\phi\rightarrow \eta'\gamma}$ & $0.265\pm 0.012$   &  $0.265\pm0.027$  \\
$\Gamma_{\eta'\rightarrow\omega\gamma}$ & $6.2\pm 1.1 $ \,\,  & \,\,  $6.2\pm 0.4$  \\
$\Gamma_{\eta\rightarrow\gamma\gamma}$ & $0.510\pm 0.026 $   \,\,  & \,\, $0.463\pm 0.038$  \\
$\Gamma_{\eta'\rightarrow\gamma\gamma}$ & $4.30\pm 0.15$   \,\,  & \,\, $4.13\pm 0.26$  \\
$\Gamma_{\eta\rightarrow\gamma e^- e^+}$ & $(8.8\pm 1.6)\times 10^{-3}$   \,\,  & \,\, $(7.7\pm 0.6)\times 10^{-3}$  \\
$\Gamma_{\eta\rightarrow\gamma \mu^- \mu^+}$ & $(0.40\pm 0.08)\times 10^{-3}$   \,\,  & \,\, $(0.36\pm 0.03)\times 10^{-3}$  \\
$\Gamma_{\eta'\rightarrow\gamma \mu^- \mu^+}$ & $(2.1\pm 0.7)\times 10^{-2}$   \,\,  & \,\,$(1.6\pm 0.1)\times 10^{-2}$  \\
$\Gamma_{\omega \rightarrow\pi e^- e^+}$ &   $6.54\pm0.83$  & $6.81\pm0.30$  \\
$\Gamma_{\omega \rightarrow\pi \mu^- \mu^+}$ & $0.82\pm0.21$  & $0.67\pm0.03$ \\
$\Gamma_{\phi \rightarrow\eta e^- e^+}$ & $ 0.490\pm 0.048$  \,\,  & \,\,  $0.464\pm 0.031$  \\
\hline\hline
\end{tabular}
\caption{\label{table-NoJpsidecay}  Experimental and theoretical
values of the decay widths of various processes: $P\rightarrow
V\gamma$, $V\rightarrow P\gamma$, $P\rightarrow \gamma\gamma$,
$P\rightarrow \gamma l^{+}l^{-}$, $V\rightarrow Pl^{+}l^{-}$. The
experimental data are taken from~\cite{Pdg}. All of the values are
given in units of KeV. The error bands of the theoretical outputs
 are calculated by using the parameter configurations in Talbe~\ref{tablepar}.
  }
\end{center}
\end{small}
\end{table}

Some remarks for the fitting results are in order. We comment them one by one as follows.

\begin{enumerate}
\item  The $\eta-\eta'$ mixing parameters. In the $U(3)$ $\chi$PT, $F_8$ can be fixed through the ratio of
$F_K/F_\pi$ at the next-to-next-to-leading order within the triple
expansion scheme, i.e. a simultaneous expansion on the momentum,
quark mass and $1/N_C$. This approach leads to the prediction
$F_8 = 1.34 F_\pi$~\cite{Kaiser98}. While for $F_0$, according to
the results from our previous work with only light
hadrons~\cite{chen2012}, its error bar is much larger than that of $F_8$.
After including the $J/\psi$ data, we find that the error bar of
$F_0$ is now compatible with the one for $F_8$, indicating the
sensitivity of this parameter in $J/\psi$ decays. $F_0$ was
determined in the process $P\to\gamma\gamma$ at next-to-leading
order by ignoring the chiral symmetry breaking operators in
Refs.~\cite{Leutwyler98,Feldmann99}, which led to $F_0 = 1.25
F_\pi$. As one can see the numbers
in Table~\ref{tablepar}, our result for $F_8$ is slightly larger
than the $\chi$PT prediction, and our $F_0$ agrees with the $\chi$PT
prediction. About the mixing angles, our present determinations for
$\theta_8$ and $\theta_0$ are somewhat more negative than those in
literature, see Table 1 of Ref.~\cite{Feldmann00}. Comparing with
our previous determinations in Ref.~\cite{chen2012} with only light
hadron data, the present values for the two angles also become more
negative, see the last two columns in Table~\ref{tablepar}. This
tells us that the $J/\psi$ data prefer somewhat more negative mixing
angles. Nevertheless, when taking into account the errors of these
two parameters as shown in Table~\ref{tablepar}, the results of
$\theta_8$ and $\theta_0$ in this analysis are still comparable with
previous studies. It is clear that the present error bands of $\theta_0$
and $\theta_8$ are much smaller than the values in Ref.~\cite{chen2012},
which highlights the relevance of the $J/\psi$ data in the determination of
the $\eta$-$\eta'$ mixing parameters.

For the mixing parameters in the
quark-flavor basis defined in Eq.~\eqref{twoanglesmixingqs},
the theoretical prediction for the difference between the angles $\phi_q$ and $\phi_s$ should be very small, since their difference
is caused by the OZI-rule violating terms.
In this work, we further confirm this prediction and the difference between $\phi_q$ and $\phi_s$ is indeed found to be tiny. Our results,
$\phi_q = (34.5\pm 1.8 )^{\circ}$ and $\phi_s = (36.0\pm 1.4)^{\circ}$, are in qualitative agreement with these earlier studies
in Refs.~\cite{Feldmann00,Thomas}, which give the result around $40^{\circ}$. Our analysis prefers slightly smaller magnitudes of $\phi_q$ and $\phi_s$.

\item  The $\tilde{c}_3$ and $\tilde{d}_2$ parameters were determined with huge
error bars in our previous study without the $J/\psi$ data~\cite{chen2012}.
We see that the present results are compatible with those in~\cite{chen2012}, but have
smaller error bars now. The magnitude of $\tilde{c}_3$ is of order $10^{-3}$ now, which
is consistent with the magnitudes of $\tilde{c}_4$ and $\tilde{c}_6$ determined in Ref.~\cite{chen2014}.

\item  The $J/\psi \to P l^+ l^-$ process. In Ref.~\cite{HBLi2012}, the form
factor of $J/\psi \to P \gamma^{\ast}$ is parameterized by using the
simple pole approximation in the VMD framework  as
\begin{eqnarray}
F_{\psi P}(s)\equiv \frac{G_{\psi\rightarrow
P\gamma^{\ast}}(s)}{G_{\psi\rightarrow P\gamma^{\ast}}(0)}
=\frac{1}{1-s/\Lambda^2}, \label{eq-ffpsiP-VMD}
\end{eqnarray}
 with $\Lambda$ chosen to be the mass
of $\psi^{\prime}$.  We plot the
integrand of Eq.~\eqref{gammapsipll}, namely the differential decay
widths for $J/\psi \to P l^+ l^-$ in
Fig.~\ref{fig.JpsiPllDifDecayWidths}, from which it is not difficult to observe that the $J/\psi \to P l^+ l^-$ decay width is
dominated by the small $s$ region, purely due to
the kinematic factors. Therefore the $s/\Lambda^2$ term in
Eq.~\eqref{eq-ffpsiP-VMD} cannot give large effect. This also
explains that different values of $\beta$ in the form factor
$e^{\frac{s}{16\beta^2}}$ in Eq.~\eqref{Eqpsietacg} make little
difference. Thus for the $J/\psi \to P l^+ l^-$ decay rate, the
overwhelmingly dominant part is from the structure independent
factor $F_{\psi P}(s)=1$ and any model-dependent hadronic
corrections to $F_{\psi P}(s)=1$ will only slightly affect the total
decay rate.

\begin{figure}[ht]
\includegraphics[height=8cm,width=12cm]{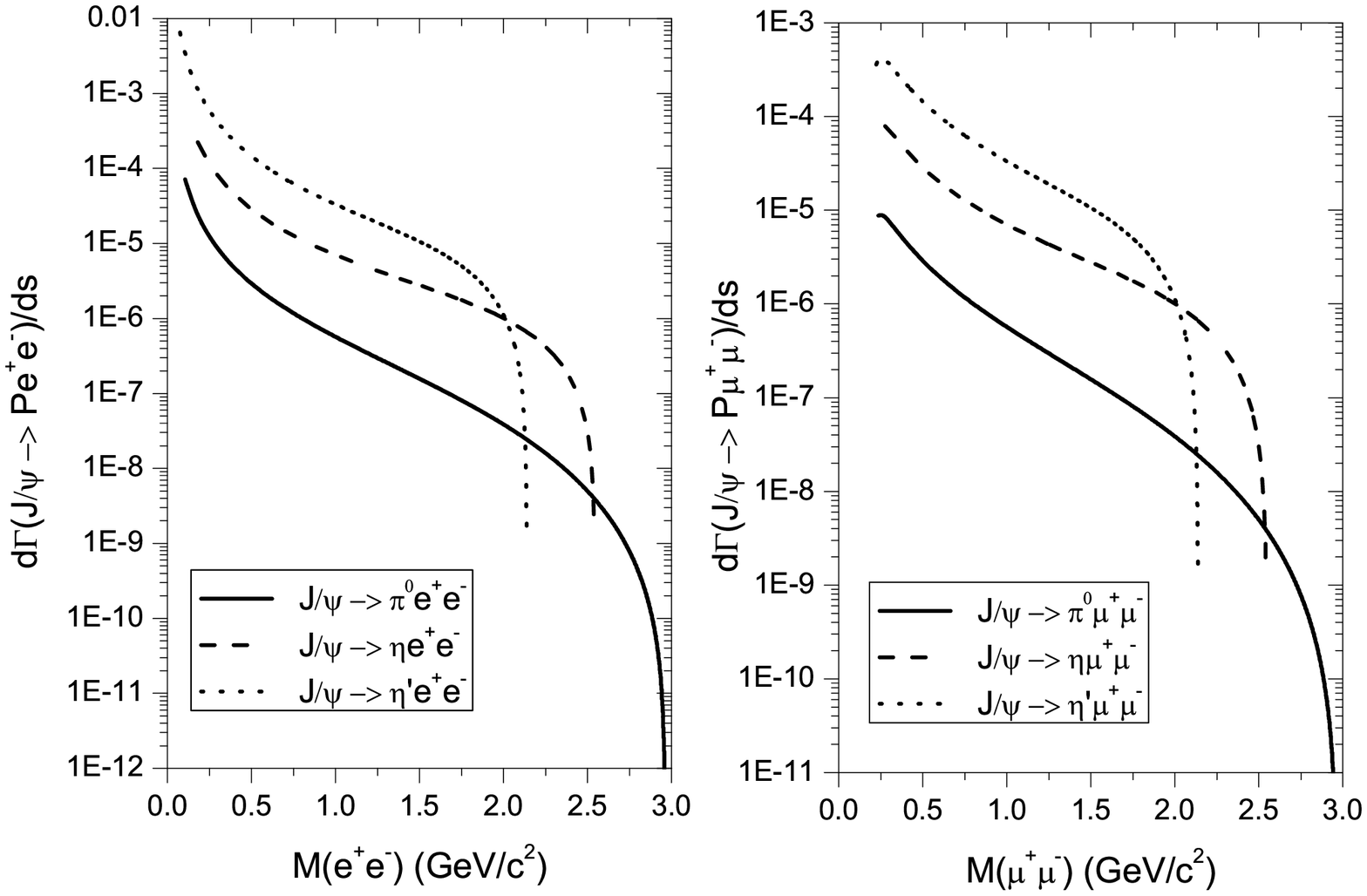}
\caption{ The differential decay widths of $J/\psi \to P l^+ l^-$
processes, where the solid line is for $J/\psi \to \pi^0 l^+ l^-$,
the dashed line for $J/\psi \to \eta l^+ l^-$, and the dotted line
for $J/\psi \to \eta^\prime l^+ l^-$. The left panel is for the
lepton pair $e^+e^-$ and the right panel for the lepton pair
$\mu^+\mu^-$. \label{fig.JpsiPllDifDecayWidths} }
\end{figure}

At a first glance, the theoretical model we propose to study the $J/\psi \to P l^+ l^-$ decay, which is schematically depicted in Fig.~\ref{fig.psipg},
is clearly different from the VMD model in Eq.~\eqref{eq-ffpsiP-VMD}, since we do not explicitly include the effects of $\psi'$ in Fig.~\ref{fig.psipg}.
Nevertheless, as discussed just before, what matters to the decay rate of $J/\psi \to P l^+ l^-$ is the very low energy region of the
integrand in Eq.~\eqref{gammapsipll}, where the propagator of the $\psi'$ from the VMD approach in Eq.~\eqref{eq-ffpsiP-VMD} reduces to a constant.
Essentially, the diagrams (a) and (c) in Fig.~\ref{fig.psipg} give constant terms in the low energy region.
While the diagram (b) has to be subtracted in order to be consistent with experimental setup,
as the contributions from the light vector resonances have been removed in the final results from experimental analyses~\cite{BESIII2014}.
Therefore we can conclude that our model in Fig.~\ref{fig.psipg} is qualitatively similar as the commonly used VMD model in Eq.~\eqref{eq-ffpsiP-VMD}
when focusing on the $J/\psi \to P l^+ l^-$ decay width.
The theoretical outputs and the experimental data of the $J/\psi \to P l^+ l^-$ processes are
summarized in Table~\ref{table-JpsitoPll}.

Both our theoretical outputs and the VMD predictions for the $J/\psi
\to\eta(\eta^{\prime}) e^+ e^-$ processes agree with the data, as
shown in Table~\ref{table-JpsitoPll}. While for the $J/\psi \to\pi^0
e^+ e^-$ process, none of the results from the two approaches are
compatible with the experimental data. Notice that for the
experimental analyses of the $J/\psi \to P e^+ e^-$ decays in
Ref.~\cite{BESIII2014}, the peaking backgrounds from the
intermediate processes like $J/\psi \to \rho^0 P, \omega P$ and
$\phi P$, with $\rho^0, \omega$ and $\phi$ decaying into $e^+e^-$,
have been subtracted. From the world average results in
Ref.~\cite{Pdg}, we know the branching ratios $B_{J/\psi
\rightarrow\rho^0\pi^0}=(5.6\pm 0.7)\times 10^{-3}$ and $B_{\rho^0
\rightarrow e^+e^-}=(4.72\pm 0.05)\times 10^{-5}$, so that
$B_{J/\psi \rightarrow \rho^0\pi^0} \times B_{\rho^0 \rightarrow
e^+e^-}=(2.64\pm 0.36)\times 10^{-7}$, which is about $1/3$ of the
branching ratio $B_{J/\psi \rightarrow \pi e^+ e^-}$ given in
Ref.~\cite{BESIII2014}. This rough estimate tells us that the
contributions from the intermediate processes with light hadrons can be
important and this conclusion is in accord with the dispersive 
analyses in Ref.~\cite{Kubis:2014gka}. Our simple estimate also confirms the findings in
Refs.~\cite{Rosner,QZhao2011}, where the dominance of
$J/\psi\rightarrow \pi^0\rho^0 \rightarrow \pi^0\gamma$ in the
$J/\psi \rightarrow \pi^0\gamma$ decay is evident.
In our theoretical scheme, we find
large destructive interference between the intermediate-$\rho^0$'s
contribution and other contributions in the $J/\psi \rightarrow
\pi^0\gamma$ and the $J/\psi \rightarrow \pi^0 l^+l^-$ processes, so that
neglecting the intermediate-$\rho^0$'s contribution leads to a larger
value of the branching ratio of $J/\psi \rightarrow \pi^0 e^+e^-$.

Therefore we urge the experimental colleagues to take more serious analyses of
the light vector contributions in the $J/\psi \to \pi^0 e^+ e^-$ decays in order
to clarify its decay mechanism. We find that the contributions from the intermediate light
vectors are tiny in the $J/\psi\rightarrow
\eta(\eta^{\prime})\gamma$ and $J/\psi \rightarrow
\eta(\eta^{\prime})l^+l^-$ processes.

In addition, we provide the predictions for the $J/\psi \to P \mu^+
\mu^-$ decays in Table~\ref{table-JpsitoPll} together with the
results from Ref.~\cite{HBLi2012}. We hope our results can provide
to the experimental colleagues some references for future measurements
on these channels.

\begin{table}[h]
\begin{small}
\begin{center}
\begin{tabular}{c|c|c|c}\hline\hline
&  Exp. data & this work    & VMD prediction~\cite{HBLi2012}      \\
\hline
$\psi \to \pi^0 e^+ e^-$ &$0.0756\pm 0.0141$ & $ 0.1191\pm 0.0138 $      &$0.0389_{-0.0033}^{+0.0037} $   \\
$\psi \to \eta e^+ e^-$  & $1.16\pm 0.09$ &  $ 1.16 \pm 0.08  $     &$1.21\pm 0.04$  \\
$\psi \to \eta^{'}e^+ e^-$  &$5.81\pm 0.35$ & $ 5.76\pm 0.16 $ &$5.66\pm 0.16$ \\
$\psi \to \pi^0 \mu^+ \mu^-$  & - & $ 0.0280\pm 0.0032 $      &$0.0101_{-0.0009}^{+0.0010} $  \\
$\psi \to \eta \mu^+ \mu^-$ & - &  $ 0.32 \pm 0.02 $     &$0.30\pm 0.01$   \\
$\psi \to \eta^{'}\mu^+ \mu^-$  & -& $ 1.46 \pm 0.04$    &$1.31\pm 0.04$   \\
\hline\hline
\end{tabular}
\caption{\label{table-JpsitoPll} Branching ratios ($\times 10^{-5}$) for $J/\psi \to P
l^+ l^-$, where $P=\pi^0,\eta,\eta^{'},$ and
$\l=e,\mu$. }\end{center}
\end{small}
\end{table}

\item $\eta(\eta^{'})-\eta_c$ mixing.
For the $J/\psi \to\eta(\eta^{\prime}) \gamma^{(*)}$ processes, if we do not include the mechanism raised in Ref.~\cite{KTChao1990},
i.e. the diagram (c) in Fig.~\ref{fig.psipg}, there is no way for us to simultaneously
describe the $J/\psi \to\eta(\eta^{\prime}) \gamma^{(*)}$ processes together with other types of data. Therefore
we confirm the importance of the $\eta(\eta')$-$\eta_c$ mixing in $J/\psi \to\eta(\eta^{\prime}) \gamma^{(*)}$
as advocated in Ref.~\cite{KTChao1990,QZhao2011}.
In Table~\ref{table-etacmixing}, we quantitatively show the contributions from the diagram (c) in Fig.~\ref{fig.psipg} to the total decay widths
of $J/\psi \to\eta(\eta^{\prime}) \gamma$ and $J/\psi \to\eta(\eta^{\prime}) e^+ e^-$.

\begin{table}[h]
\begin{small}
\begin{center}
\begin{tabular}{c|c|c|c}\hline\hline
&     Exp. data & $\eta_c$ mixing in this work&   $\eta_c$ mixing in Ref.~\cite{QZhao2011}       \\
\hline
$\psi \to \eta\gamma$ & $1.104\pm 0.034$ & 0.823   &0.61   \\
$\psi \to \eta^{'}\gamma$& $5.16\pm 0.15$ & 4.56   &3.5   \\
$\psi \to \eta e^+ e^-$ &$(1.16\pm 0.09)\times 10^{-2}$ & $0.95\times 10^{-2}$   & -   \\
$\psi \to \eta^{'}e^+ e^-$ & $(5.81\pm
0.35)\times 10^{-2}$ & $5.07\times 10^{-2}$  & -  \\
\hline\hline
\end{tabular}
\caption{\label{table-etacmixing}  Branching ratios ($\times 10^{-3}$) for $J/\psi \to
\eta(\eta^{'})\gamma$, and $J/\psi \to \eta(\eta^{'})e^+ e^-$
caused by the $\eta(\eta^{'})-\eta_c$ mixing.   }
\end{center}
\end{small}
\end{table}

\item The roles of the EM and strong transitions in the $J/\psi \to VP$ decays.
In order to discuss the interplay roles of the EM and strong interactions in the $J/\psi \to VP$
processes, we show the modulus of the form factors $G_{VP}$ defined in Eq.~\eqref{defgvp} in Table~\ref{table-EMinJpsiVP}.
In the left side of this table we show the contributions from strong interactions to the isospin conserved channels
and in the right side we show the EM contributions to the isospin violated channels.
It is clear that the strong interactions play the dominant roles in the isospin conserved decay channels
and the EM interactions dominate the isospin violated channels.
Furthermore, for the isospin conserved cases we have explicitly checked that there are no significant contributions from
the EM transitions, with the exception of the $J/\psi \rightarrow \phi\eta^{\prime}$ channel.
We find that there is a large destructive interference between the strong and EM interactions in this process.
Generally speaking, our findings in $J/\psi \to VP$ decays are consistent with a general expectation
and our numbers are in qualitative agreement with those in Ref.~\cite{QZhao2007}.

\begin{table}[h]
\begin{small}
\begin{center}
\begin{tabular}{c|c|c| |c|c|c}\hline\hline
 Isospin conserved cases & Exp. data & Strong interaction  &Isospin violated cases& Exp. data & EM interaction        \\
\hline
$\big|G_{\psi \to \rho^0\pi^0}\big|$ & $2.541\pm0.154$ & $ 2.933\pm 0.144$ &$\big|G_{\psi \to \rho^0\eta}\big|$ & $0.498\pm0.029$ &   $0.510\pm 0.056$\\
$\big|G_{\psi \to \rho\pi}\big|$ & $4.415\pm0.192$  &$ 5.080\pm 0.250$ &$\big|G_{\psi \to \rho^0\eta^{'}}\big|$& $0.418\pm0.034$  &  $ 0.429 \pm0.063$ \\
$\big|G_{\psi \to \omega\eta}\big|$ &$1.499\pm 0.084$  &$ 1.628\pm 0.097$ &$\big|G_{\psi \to \omega\pi^0}\big|$ & $0.722\pm0.039$ & $0.722 \pm 0.091$ \\
$\big|G_{\psi \to \omega\eta^{'}}\big|$ & $0.552\pm 0.031$ & $ 0.659\pm 0.059$ &&  \\
$\big|G_{\psi \to \phi\eta}\big|$ &  $1.069\pm 0.056$ & $ 1.346\pm 0.066$  &&\\
$\big|G_{\psi \to \phi\eta^{'}}\big|$ &  $0.910\pm 0.076$ &$ 1.178\pm 0.126$  &&\\
$\big|G_{\psi \to K^{\ast +} K^-}\big|$ &  $1.860\pm 0.054 $ & $ 2.473\pm 0.089$  &&\\
$\big|G_{\psi \to K^{\ast 0} \bar{K}^0}\big|$ &  $1.726\pm 0.060 $ &$ 2.468\pm 0.082$ && \\
\hline\hline
\end{tabular}
\caption{\label{table-EMinJpsiVP}  The modulus of form factor
$\big|G_{\psi \to V P}\big|$ in units of $10^{-6}$MeV$^{-1}$
contributed by the strong transitions to the isospin conserved
channels, and by the EM transitions to the isospin violated
channels. The error bands from this table are calculated by using
the same parameter configurations as in Table~\ref{tablepar}. }
\end{center}
\end{small}
\end{table}

\end{enumerate}

\section{Conclusions}
\label{conclu}

We use the effective Lagrangian approach to simultaneously study the
decays of $J/\psi \to VP$, $J/\psi \to P \gamma$, $J/\psi \to P l^+
l^-$ together with the light meson radiative processes, such as
$VP\gamma^{(*)}$, $P\gamma\gamma^{(*)}$. We take the building blocks
involving external sources, light pseudoscalar mesons and vector resonances from the
chiral effective field theory to construct the effective Lagrangian
for the $J/\psi$ decays. The $SU(3)$-flavor symmetry breaking
effects and the OZI rules are systematically and concisely
implemented in this approach. For the processes with only light
hadrons, we follow closely our previous work in Ref.~\cite{chen2012}
and use the resonance chiral theory to build the relevant
Lagrangians. Two-mixing-angle scheme from the general discussion in
$U(3)$ chiral perturbation theory is employed to describe the
$\eta$-$\eta'$ mixing in various processes involving $\eta$ or
$\eta'$. Comparing with our previous results by taking only the
light hadron data in the analyses, we update the values for the
mixing parameters by including the relevant $J/\psi$ decays in this
work. It turns out that the present determination prefers more
negative values for the two mixing angles $\theta_0$ and $\theta_8$
from the octet-singlet basis, or smaller values for the $\phi_q$ and
$\phi_s$ from the quark-flavor basis. Since we make a global fit for
the $J/\psi$ and the light hadron data in this work, smaller error
bars result for some of the couplings in Table~\ref{tablepar},
especially for $\tilde{c}_3$ and $\theta_0$. This clearly indicates
the relevance of the $J/\psi$ data for the determinations of the
couplings involving only light hadrons and the $\eta$-$\eta'$ mixing
parameters.

In a short summary, we have found a proper theoretical framework that can be used to
systematically and successfully describe the $J/\psi \to VP$, $P\gamma^{(*)}$
and the light meson radiative decays. Another interesting and relevant subjects along this research line is to
take the $\psi(2S)\to V P$ and $\psi(2S) \to P\gamma$ into account, so that the famous $\rho\pi$ puzzle in charmonium
decays could be addressed. Nevertheless, a straightforward generalization of the decay mechanisms from $J/\psi$ 
to $\psi(2S)$ might be problematic, as recently discussed in Ref.~\cite{Gerard:2013gya}. 
Moreover, due to the fact that the low statistics for the $\psi(2S)$ data can not
be compared to the precise ones of the $J/\psi$ and light hadrons, it is not so clear whether it is 
justified to make a global fit by including the $\psi(2S)$ data as the ones considered in this work.
Therefore we think it is worthy starting an independent project to study the $\psi(2S)$ decays, specially to address the long standing $\rho\pi$ puzzle, which
is under preparation.

\section*{Acknowledgments}
We acknowledge Xin-Kun Chu, Jian-Ping Dai, and Rong-Gang Ping from BESIII collaboration for
helpful discussions on the experimental analyses. This work is
supported in part by the National Natural Science Foundation of
China under Grants No.~11105038, No.~11035006, No.~11121092, and No.~11261130311
(CRC110 by DFG and NSFC), the Chinese Academy of Sciences under
Project No.~KJCX2-EW-N01, and the Ministry of Science and Technology
of China (2009CB825200). Z.H.G. acknowledges the grants from the
Education Department of Hebei Province under Contract No.~YQ2014034, 
the grants from the Department of Human Resources and Social Security of Hebei Province with contract No.~C201400323, 
and the Doctor Foundation of Hebei Normal University under Contract
No.~L2010B04.

\appendix

\section{ The form factors of $J/\psi \rightarrow P\gamma^{\ast}$ }\label{ffpsipg}

 For latter convenience, we define several $a_i$ factors as follows
 \begin{eqnarray}
a_1 = && \frac{F}{\cos(\theta_0-\theta_8)}\left(\frac{1}{\sqrt{6}}\frac{\cos \theta_0}{F_8}-\frac{1}{\sqrt{3}}\frac{\sin \theta_8}{F_0}\right)
\,, \nn \\
a_2 = && \frac{F}{\cos(\theta_0-\theta_8)}\left(\frac{1}{\sqrt{6}}\frac{\sin \theta_0}{F_8}+\frac{1}{\sqrt{3}}\frac{\cos \theta_8}{F_0}\right)
\,, \nn \\
a_3 = && \frac{F}{\cos(\theta_0-\theta_8)}\left(-\frac{2}{\sqrt{6}}\frac{\cos \theta_0}{F_8}-\frac{1}{\sqrt{3}}\frac{\sin \theta_8}{F_0}\right)
\,, \nn \\
a_4 = && \frac{F}{\cos(\theta_0-\theta_8)}\left(-\frac{2}{\sqrt{6}}\frac{\sin \theta_0}{F_8}+\frac{1}{\sqrt{3}}\frac{\cos \theta_8}{F_0}\right)
\nn\,.
\end{eqnarray}

The explicit expressions for the form factors of the $J/\psi \to P\gamma^{\ast}$ defined
in Eq.~\eqref{defgpg} are given below:

\begin{eqnarray}
G_{\psi \to \pi^0
\gamma^\ast}(s)=-\frac{4g_1}{F_\pi}-\frac{16g_2}{F_\pi
M_\psi^2}m_\pi^2 +2\sqrt{2} h_1 M_\psi \frac{F_V}{F_\pi} D_\rho(s)
+8\sqrt{2} h_2 \frac{m_\pi^2}{M_\psi} \frac{F_V}{F_\pi} D_\rho(s)\,,
\end{eqnarray}

\begin{eqnarray}
G_{\psi \to \eta
\gamma^\ast}(s)=&&-\frac{4\sqrt{2}g_1}{3F}(a_1-a_3)-\frac{16\sqrt{2}g_2}{3F
M_\psi^2}\big[a_1 m_\pi^2-a_3(2m_K^2-m_\pi^2)\big]+\frac{4}{3}
M_\psi \frac{F_V}{F}\big[(h_1 +4h_2\frac{m_\pi^2}{M_\psi^2}+2h_3)a_1
\nonumber\\&&+h_3 a_3\big]D_\omega(s)-\frac{4}{3} M_\psi
\frac{F_V}{F}\bigg\{\big[h_1
+4h_2\frac{1}{M_\psi^2}(2m_K^2-m_\pi^2)+h_3\big]a_3 +2h_3
a_1\bigg\}D_\phi(s)
\nonumber\\&&+\lambda_{\eta\eta_c}g_{\psi\eta_c\gamma}(s)e^{i\delta_\eta}\,,
\end{eqnarray}

\begin{eqnarray}
G_{\psi \to \eta^{'}
\gamma^\ast}(s)=&&-\frac{4\sqrt{2}g_1}{3F}(a_2-a_4)-\frac{16\sqrt{2}g_2}{3F
M_\psi^2}\big[a_2 m_\pi^2-a_4(2m_K^2-m_\pi^2)\big]+\frac{4}{3}
M_\psi \frac{F_V}{F}\big[(h_1 +4h_2\frac{m_\pi^2}{M_\psi^2}+2h_3)a_2
\nonumber\\&&+h_3 a_4\big]D_\omega(s)-\frac{4}{3} M_\psi
\frac{F_V}{F}\bigg\{\big[h_1
+4h_2\frac{1}{M_\psi^2}(2m_K^2-m_\pi^2)+h_3\big]a_4 +2h_3
a_2\bigg\}D_\phi(s)
\nonumber\\&&+\lambda_{\eta^{\prime}\eta_c}g_{\psi\eta_c\gamma}(s)e^{i\delta_{\eta^{\prime}}}\,,
\end{eqnarray}
where the definition of $D_R(s)$ is
\begin{eqnarray}\label{defdr}
D_R(s) = \frac{1}{M_R^2 -s - i M_R \Gamma_R(s) } \,.
\end{eqnarray}
For the narrow-width resonances $\omega$ and $\phi$,  we use the constant
widths in the numerical discussion. For the $\rho$ resonance,
the energy dependent width is given by~\cite{chen2012}
\begin{eqnarray}
\Gamma_{\rho}(s)=\frac{sM_V}{96\pi
F^2}[\sigma_{\pi}^3\theta(s-4m_{\pi}^2)+\frac{1}{2}\sigma_K^3\theta(s-4m_K^2)]\,,
\end{eqnarray}
where $\sigma_P=\sqrt{1-4m_P^2/s}$ and $\theta(s)$ is the step
function.

\section{ The form factors of $VP\gamma^{*}$ }\label{ffvpg}

The various form factors $F_{VP\gamma^{*}}(s)$ from different
processes have already been given in Ref.~\cite{chen2012} and we show them below for the sake of completeness:

\begin{eqnarray}
F_{\rho \pi \gamma^{\ast}}(s)=&&  -\frac{2\sqrt2 }{3F_\pi M_V
M_\rho} \big[
(\tilde{c}_1+\tilde{c}_2+8\tilde{c}_3-\tilde{c}_5)m_\pi^2
+(\tilde{c}_2+\tilde{c}_5
-\tilde{c}_1-2\tilde{c}_6)M_\rho^2+(\tilde{c}_1-\tilde{c}_2+\tilde{c}_5)s
\big] \nonumber
\\&&  +\frac{4 F_V}{3F_\pi M_\rho }D_\omega(s) \big[
(\tilde{d}_1+8\tilde{d}_2-\tilde{d}_3)m_\pi^2 +\tilde{d}_3
(M_\rho^2+s) \big]  \,,
\end{eqnarray}

\begin{eqnarray}
F_{\rho\eta\gamma^\ast}(s)=&&  -\frac{4}{ M_V M_\rho F}a_1\big[
M_\rho^2(\tilde{c}_2-\tilde{c}_1+\tilde{c}_5-2\tilde{c}_6)+m_\eta^2(\tilde{c}_2+\tilde{c}_1-\tilde{c}_5)
 +8 \tilde{c}_3 m_\pi^2 \nonumber\\&&
+(\tilde{c}_1-\tilde{c}_2+\tilde{c}_5)s \big]  + \frac{4\sqrt2 F_V}{
M_\rho F}D_\rho(s)a_1\big[
\tilde{d}_3(M_\rho^2-m_\eta^2+s)+\tilde{d}_1m_\eta^2+8 \tilde{d}_2
m_\pi^2 \big]\nonumber \\
&& -\frac{\sin\theta_8}{\cos{(\theta_0-\theta_8)}F_0} \big[
-\frac{4\sqrt2 M_V}{ M_\rho} \tilde{c}_8+ \frac{8 F_V M_V^2}{
M_\rho} \tilde{d}_5 D_\rho(s) \big]  \,,
\end{eqnarray}

\begin{eqnarray}
F_{\rho\eta'\gamma^\ast}(s)= && -\frac{4}{ M_V M_\rho F} a_2
 \big[ M_\rho^2(\tilde{c}_2-\tilde{c}_1+\tilde{c}_5-2\tilde{c}_6)+m_{\eta'}^2(\tilde{c}_2+\tilde{c}_1-\tilde{c}_5)
  +8 \tilde{c}_3 m_\pi^2 \nonumber \\ &&
 +(\tilde{c}_1-\tilde{c}_2+\tilde{c}_5)s \big]
 + \frac{4\sqrt2 F_V}{ M_\rho F} D_\rho(s) a_2 \big[
\tilde{d}_3(M_\rho^2-m_{\eta'}^2+s)+\tilde{d}_1m_{\eta'}^2+8
\tilde{d}_2 m_\pi^2 \big] \nonumber \\
&& - \frac{\cos\theta_8}{\cos{(\theta_0-\theta_8)}F_0}  \big[
\frac{4\sqrt2 M_V}{ M_\rho} \tilde{c}_8-\frac{8F_V M_V^2}{ M_\rho}
\tilde{d}_5 D_\rho(s) \big]\,,
\end{eqnarray}

\begin{eqnarray}
F_{\omega\pi\gamma^\ast}(s) = && -\frac{2\sqrt2 }{F_\pi M_V
M_\omega} \big[
(\tilde{c}_1+\tilde{c}_2+8\tilde{c}_3-\tilde{c}_5)m_\pi^2
+(\tilde{c}_2+\tilde{c}_5
-\tilde{c}_1-2\tilde{c}_6)M_\omega^2+(\tilde{c}_1-\tilde{c}_2+\tilde{c}_5)s
\big] \nonumber \\&&  +\frac{4 F_V}{F_\pi M_\omega} D_\rho(s) \big[
(\tilde{d}_1+8\tilde{d}_2-\tilde{d}_3)m_\pi^2 +\tilde{d}_3
(M_\omega^2+s)  \big]+r_1 M_\omega D_{\rho^{\prime}}(s) \,,
\end{eqnarray}

\begin{eqnarray}
F_{\omega\eta\gamma^{\ast}}(s) =&& -\frac{4}{3 M_V M_\omega
F}a_1\big[
M_\omega^2(\tilde{c}_2-\tilde{c}_1+\tilde{c}_5-2\tilde{c}_6)
+m_\eta^2(\tilde{c}_2+\tilde{c}_1-\tilde{c}_5) +8 \tilde{c}_3
m_\pi^2 \nonumber
\\ &&+(\tilde{c}_1-\tilde{c}_2+\tilde{c}_5)s \big]  +
\frac{4\sqrt{2} F_V}{3 M_\omega F} D_\omega(s) a_1 \big[
\tilde{d}_3(M_\omega^2-m_\eta^2+s)+\tilde{d}_1m_\eta^2+8 \tilde{d}_2
m_\pi^2 \big]\nonumber \\
&&-\frac{\sin\theta_8}{3 \cos{(\theta_0-\theta_8)} F_0} \big[
-\frac{4\sqrt2 M_V}{ M_\omega} \tilde{c}_8+ \frac{8 F_V M_V^2}{
M_\omega} \tilde{d}_5 D_\omega(s) \big]  \,,
\end{eqnarray}

\begin{eqnarray}
F_{\omega\eta'\gamma^\ast}(s) =&& -\frac{4}{3 M_V M_\omega F} a_2
 \big[ M_\omega^2(\tilde{c}_2-\tilde{c}_1+\tilde{c}_5-2\tilde{c}_6)+m_{\eta'}^2(\tilde{c}_2+\tilde{c}_1-\tilde{c}_5)
  +8 \tilde{c}_3 m_\pi^2 \nonumber \\ &&
 +(\tilde{c}_1-\tilde{c}_2+\tilde{c}_5)s \big]
 + \frac{4\sqrt{2} F_V}{3 M_\omega F} D_\omega(s) a_2 \big[
\tilde{d}_3(M_\omega^2-m_{\eta'}^2+s)+\tilde{d}_1m_{\eta'}^2+8
\tilde{d}_2 m_\pi^2 \big]\nonumber \\
&& -\frac{\cos\theta_8}{3 \cos{(\theta_0-\theta_8)} F_0}  \big[
\frac{4\sqrt2 M_V}{ M_\omega} \tilde{c}_8-\frac{8F_V M_V^2}{
M_\omega} \tilde{d}_5 D_\omega(s) \big] \bigg\}\,,
\end{eqnarray}

\begin{eqnarray}
F_{\phi\eta\gamma^\ast}(s)=&&  -\frac{4\sqrt2}{3 M_V M_\phi F} a_3
 \big[ M_\phi^2(\tilde{c}_2-\tilde{c}_1+\tilde{c}_5-2\tilde{c}_6)+m_\eta^2(\tilde{c}_2+\tilde{c}_1-\tilde{c}_5)
  +8 \tilde{c}_3 (2m_K^2-m_\pi^2)\nonumber \\ &&
 +(\tilde{c}_1-\tilde{c}_2+\tilde{c}_5)s \big]
 + \frac{8 F_V}{3 M_\phi F} D_\phi(s)a_3 \big[
\tilde{d}_3(M_\phi^2-m_\eta^2+s)+\tilde{d}_1m_\eta^2+8
\tilde{d}_2(2m_K^2-m_\pi^2) \big]\nonumber \\
&& -\frac{\sin\theta_8}{\cos{(\theta_0-\theta_8)}F_0}  \big[
-\frac{8 M_V}{3 M_\phi} \tilde{c}_8+ \frac{8\sqrt2 F_V M_V^2}{3
M_\phi} \tilde{d}_5 D_\phi(s) \big] \,,
\end{eqnarray}

\begin{eqnarray}
F_{\phi\eta'\gamma^\ast}(s)=&&  -\frac{4\sqrt2}{3 M_V M_\phi F} a_4
 \big[ M_\phi^2(\tilde{c}_2-\tilde{c}_1+\tilde{c}_5-2\tilde{c}_6)+m_{\eta'}^2(\tilde{c}_2+\tilde{c}_1-\tilde{c}_5)
  +8 \tilde{c}_3 (2m_K^2-m_\pi^2)\nonumber \\ &&
 +(\tilde{c}_1-\tilde{c}_2+\tilde{c}_5)s \big]
 + \frac{8 F_V}{3 M_\phi F} D_\phi(s) a_4 \big[
\tilde{d}_3(M_\phi^2-m_{\eta'}^2+s)+\tilde{d}_1m_{\eta'}^2+8
\tilde{d}_2(2m_K^2-m_\pi^2) \big]\nonumber \\
&& + \frac{\cos\theta_8}{\cos{(\theta_0-\theta_8)}F_0}  \big[
-\frac{8 M_V}{3 M_\phi} \tilde{c}_8+ \frac{8\sqrt2 F_V M_V^2}{3
M_\phi} \tilde{d}_5 D_\phi(s) \big] \,,
\end{eqnarray}

\begin{eqnarray}
F_{K^{\ast+}K^-\gamma^\ast}(s)=&&  -\frac{2\sqrt2 }{3F_K M_V
M_{K^*}} \big[
(\tilde{c}_1+\tilde{c}_2+8\tilde{c}_3-\tilde{c}_5)m_K^2
+(\tilde{c}_2+\tilde{c}_5
-\tilde{c}_1-2\tilde{c}_6)M_{K^*}^2+(\tilde{c}_1-\tilde{c}_2+\tilde{c}_5)s
\nonumber \\&& +24 \tilde{c}_4 (m_K^2-m_\pi^2) \big]  +\frac{2
F_V}{3F_K M_{K^*}} \big[ (\tilde{d}_1+8\tilde{d}_2-\tilde{d}_3)m_K^2
+\tilde{d}_3 (M_{K^*}^2+s) \big]\nonumber \\&&\big[ D_\omega(s) +3
D_\rho(s)-2 D_\phi(s) \big]  \,,
\end{eqnarray}

\begin{eqnarray}
F_{K^{\ast 0}\bar{K}^0\gamma^\ast}(s)= && \frac{4\sqrt2 }{3F_K M_V
M_{K^*}} \big[
(\tilde{c}_1+\tilde{c}_2+8\tilde{c}_3-\tilde{c}_5)m_K^2
+(\tilde{c}_2+\tilde{c}_5
-\tilde{c}_1-2\tilde{c}_6)M_{K^*}^2+(\tilde{c}_1-\tilde{c}_2+\tilde{c}_5)s
\big] \nonumber \\&& +\frac{2 F_V}{3F_K M_{K^*}}  \big[
(\tilde{d}_1+8\tilde{d}_2-\tilde{d}_3)m_K^2 +\tilde{d}_3
(M_{K^*}^2+s) \big]\big[ D_\omega(s) -3 D_\rho(s)-2 D_\phi(s) \big]
 \,.
\end{eqnarray}

\section{ The form factors of $J/\psi \rightarrow VP$ }\label{ffpsivp}

The explicit expressions for the form factors of the $J/\psi \to VP$ defined in Eq.~\eqref{defgvp} are given below:

\begin{eqnarray}
G_{\psi \to \rho^0\pi^0}=\frac{2\sqrt{2}}{F_\pi M_\rho}h_1 M_\psi+
\frac{8\sqrt{2}}{F_\pi M_\rho}h_2 m_\pi^2
\frac{1}{M_\psi}+\frac{32\pi\alpha}{F_\pi M_\rho}F_V
g_1+\frac{128\pi\alpha}{F_\pi M_\rho}F_V g_2
\frac{m_\pi^2}{M_\psi^2}+\frac{8\sqrt{2}\pi\alpha}{3}\frac{
f_\psi}{M_\psi}F_{\rho\pi\gamma^{*}}(M_\psi^2) \,, \nonumber \\
\end{eqnarray}

\begin{eqnarray}
G_{\psi \to \rho^+\pi^-}=\frac{2\sqrt{2}}{F_\pi M_\rho}h_1 M_\psi+
\frac{8\sqrt{2}}{F_\pi M_\rho}h_2 m_\pi^2
\frac{1}{M_\psi}+\frac{8\sqrt{2}\pi\alpha}{3}\frac{
f_\psi}{M_\psi}F_{\rho\pi\gamma^{*}}(M_\psi^2)   \,,
\end{eqnarray}

\begin{eqnarray}
G_{\psi \to \rho^0\eta}=&&\frac{32\sqrt{2}\pi\alpha}{3F M_\rho}F_V
g_1(a_1-a_3)+\frac{128\sqrt{2}\pi\alpha}{3F M_\rho M_\psi^2}F_V g_2
[a_1 m_\pi^2-a_3(2m_K^2-m_\pi^2)]\nonumber\\&&-8\pi\alpha
\frac{F_V}{M_\rho}
\lambda_{\eta\eta_c}g_{\psi\eta_c\gamma}(M_\rho^2)e^{i\delta_\eta}+\frac{8\sqrt{2}\pi\alpha}{3}\frac{
f_\psi}{M_\psi}F_{\rho\eta\gamma^{*}}(M_\psi^2) \,,
\end{eqnarray}

\begin{eqnarray}
G_{\psi \to \rho^0\eta^{'}}=&&\frac{32\sqrt{2}\pi\alpha}{3F
M_\rho}F_V g_1(a_2-a_4)+\frac{128\sqrt{2}\pi\alpha}{3F M_\rho
M_\psi^2}F_V g_2 [a_2
m_\pi^2-a_4(2m_K^2-m_\pi^2)]\nonumber\\&&-8\pi\alpha
\frac{F_V}{M_\rho}\lambda_{\eta^{\prime}\eta_c}g_{\psi\eta_c\gamma}(M_\rho^2)e^{i\delta_{\eta^{\prime}}}+\frac{8\sqrt{2}\pi\alpha}{3}\frac{
f_\psi}{M_\psi}F_{\rho\eta^{'}\gamma^{*}}(M_\psi^2) \,,
\end{eqnarray}

\begin{eqnarray}
G_{\psi \to \omega\pi^0}=\frac{32\pi\alpha}{3F_\pi M_\omega}F_V
g_1+\frac{128\pi\alpha}{3F_\pi M_\omega}F_V g_2
\frac{m_\pi^2}{M_\psi^2}+\frac{8\sqrt{2}\pi\alpha}{3}\frac{
f_\psi}{M_\psi}F_{\omega\pi\gamma^{*}}(M_\psi^2) \,,
\end{eqnarray}

\begin{eqnarray}
G_{\psi \to \omega\eta}=&&\frac{4}{F M_\omega}a_1 h_1 M_\psi+
\frac{16}{F M_\omega}a_1 h_2 m_\pi^2 \frac{1}{M_\psi}+\frac{4}{F
M_\omega}(2a_1+a_3) h_3 M_\psi+\frac{32\sqrt{2}\pi\alpha}{9F
M_\omega}F_V
g_1(a_1-a_3)\nonumber\\&&+\frac{128\sqrt{2}\pi\alpha}{9F M_\omega
M_\psi^2}F_V g_2 [a_1
m_\pi^2-a_3(2m_K^2-m_\pi^2)]-\frac{8}{3}\pi\alpha
\frac{F_V}{M_\omega}
\lambda_{\eta\eta_c}g_{\psi\eta_c\gamma}(M_\omega^2)e^{i\delta_\eta}\nonumber\\&&+\frac{8\sqrt{2}\pi\alpha}{3}\frac{
f_\psi}{M_\psi}F_{\omega\eta\gamma^{*}}(M_\psi^2)   \,,
\end{eqnarray}

\begin{eqnarray}
G_{\psi \to \omega\eta^{'}}=&&\frac{4}{F M_\omega}a_2 h_1 M_\psi+
\frac{16}{F M_\omega}a_2 h_2 m_\pi^2 \frac{1}{M_\psi}+\frac{4}{F
M_\omega}(2a_2+a_4) h_3 M_\psi+\frac{32\sqrt{2}\pi\alpha}{9F
M_\omega}F_V
g_1(a_2-a_4)\nonumber\\&&+\frac{128\sqrt{2}\pi\alpha}{9F M_\omega
M_\psi^2}F_V g_2 [a_2
m_\pi^2-a_4(2m_K^2-m_\pi^2)]-\frac{8}{3}\pi\alpha
\frac{F_V}{M_\omega}
\lambda_{\eta^{'}\eta_c}g_{\psi\eta_c\gamma}(M_\omega^2)e^{i\delta_{\eta^{'}}}\nonumber\\&&+\frac{8\sqrt{2}\pi\alpha}{3}\frac{
f_\psi}{M_\psi}F_{\omega\eta^{'}\gamma^{*}}(M_\psi^2)   \,,
\end{eqnarray}

\begin{eqnarray}
G_{\psi \to \phi\eta}=&&-\frac{2\sqrt{2}}{F M_\phi}a_3 h_1 M_\psi-
\frac{8\sqrt{2}}{F M_\phi}a_3 h_2 (2 m_K^2-m_\pi^2)
\frac{1}{M_\psi}-\frac{2\sqrt{2}}{F M_\phi}(2a_1+a_3) h_3 M_\psi
\nonumber\\&& +\frac{64\pi\alpha}{9F M_\phi}F_V
g_1(a_1-a_3)+\frac{256\pi\alpha}{9F M_\phi M_\psi^2}F_V g_2 [a_1
m_\pi^2-a_3(2m_K^2-m_\pi^2)]\nonumber\\&&-\frac{8\sqrt{2}}{3
M_\phi}\pi\alpha F_V
\lambda_{\eta\eta_c}g_{\psi\eta_c\gamma}(M_\phi^2)e^{i\delta_\eta}
+\frac{8\sqrt{2}\pi\alpha}{3}\frac{
f_\psi}{M_\psi}F_{\phi\eta\gamma^{*}}(M_\psi^2)   \,,
\end{eqnarray}

\begin{eqnarray}
G_{\psi \to \phi\eta^{'}}=&&-\frac{2\sqrt{2}}{F M_\phi}a_4 h_1
M_\psi- \frac{8\sqrt{2}}{F M_\phi}a_4 h_2 (2 m_K^2-m_\pi^2)
\frac{1}{M_\psi}-\frac{2\sqrt{2}}{F M_\phi}(2a_2+a_4) h_3 M_\psi
\nonumber\\&& +\frac{64\pi\alpha}{9F M_\phi}F_V
g_1(a_2-a_4)+\frac{256\pi\alpha}{9F M_\phi M_\psi^2}F_V g_2 [a_2
m_\pi^2-a_4(2m_K^2-m_\pi^2)]\nonumber\\&&-\frac{8\sqrt{2}}{3
M_\phi}\pi\alpha F_V
\lambda_{\eta^{'}\eta_c}g_{\psi\eta_c\gamma}(M_\phi^2)e^{i\delta_{\eta^{'}}}
+\frac{8\sqrt{2}\pi\alpha}{3}\frac{
f_\psi}{M_\psi}F_{\phi\eta^{'}\gamma^{*}}(M_\psi^2)   \,,
\end{eqnarray}

\begin{eqnarray}
G_{\psi \to K^{\ast +} K^-}=\frac{2\sqrt{2}}{F_K M_{K^{\ast}}}h_1
M_\psi+ \frac{8\sqrt{2}}{F_K M_{K^{\ast}}}h_2 m_K^2
\frac{1}{M_\psi}+\frac{8\sqrt{2}\pi\alpha}{3}\frac{
f_\psi}{M_\psi}F_{K^{\ast +} K^-\gamma^{*}}(M_\psi^2)   \,,
\end{eqnarray}

\begin{eqnarray}
G_{\psi \to K^{\ast 0} \bar{K}^0}=\frac{2\sqrt{2}}{F_K
M_{K^{\ast}}}h_1 M_\psi+ \frac{8\sqrt{2}}{F_K M_{K^{\ast}}}h_2 m_K^2
\frac{1}{M_\psi}+\frac{8\sqrt{2}\pi\alpha}{3}\frac{
f_\psi}{M_\psi}F_{K^{\ast 0} \bar{K}^0\gamma^{*}}(M_\psi^2)   \,.
\end{eqnarray}

\end{document}